\documentclass[manuscript,authorversion,screen]{acmart}
\newif\ifBlind\Blindtrue\Blindfalse
\usepackage[utf8]{inputenc}
\usepackage[T1]{fontenc}

\usepackage{booktabs} 
\usepackage{amsmath}
\usepackage{mathtools}
\usepackage{bbm}
\usepackage{subcaption}
\usepackage[ruled,linesnumbered,noend]{algorithm2e}
\usepackage{tikz}
\usepackage{enumitem}
\usepackage{array}
\usepackage[shortcuts]{extdash} 

\setcopyright{rightsretained}

\makeatletter
\@ACM@journaltrue
\@ACM@authorversiontrue
\gdef\@acmDOI{}
\gdef\@mkbibcitation{\bgroup
  \def\footnotemark{}%
  \par\medskip\small\noindent{\bfseries ACM Reference format:}\par\nobreak
  \noindent\authors. \@acmYear. \@title.
  ArXiV preprint,
  \ref{TotPages}~pages.
\par\egroup}
\makeatother

\acmYear{2017}
\copyrightyear{2017}

\newcommand{\reffig}[1]{Figure~\ref{#1}}
\newcommand{\refsec}[1]{Section~\ref{#1}}
\newcommand{\refeqn}[1]{Equation~\refeq{#1}}
\newcommand{\reftab}[1]{Table~\ref{#1}}

\newcommand{\V}[1]{#1} 
\newcommand{\pij}{\ensuremath{p_{i\kern-.5pt j}}}
\newcommand{\qij}{\ensuremath{q_{i\kern-.5pt j}}}
\newcommand{\piGj}{\ensuremath{p_{i\kern-.25pt \textit{|} \kern-.25pt j}}}
\newcommand{\pjGi}{\ensuremath{p_{\kern-.5pt j\textit{|} i}}}
\newcommand{\norm}[1]{\lVert#1\rVert}
\newcommand{\dimensional}{di\-men\-sion\-al} 

\begin{document}
\title[Semantic Word Clouds using t-SNE]{%
Semantic Word Clouds with Background Corpus Normalization
and t-distributed Stochastic Neighbor Embedding
}

\ifBlind
\author{Blind submission}
\affiliation{\vspace{4.5cm}}
\else
\author{Erich Schubert}
\orcid{0000-0001-9143-4880}
\affiliation{%
  \institution{Heidelberg University}
  \streetaddress{Im Neuenheimer Feld 205}
  \city{Heidelberg} 
  \postcode{69120}
  \country{Germany} 
}
\email{schubert@uni-heidelberg.de}

\author{Andreas Spitz}
\affiliation{%
  \institution{Heidelberg University}
  \streetaddress{Im Neuenheimer Feld 205}
  \city{Heidelberg} 
  \postcode{69120}
  \country{Germany} 
}
\email{spitz@informatik.uni-heidelberg.de}

\author{Michael Weiler}
\affiliation{%
  \institution{LMU Munich}
  \streetaddress{Oettingenstr 67}
  \city{München}
  \postcode{80538}
  \country{Germany} 
}
\email{weiler@dbs.ifi.lmu.de}

\author{Johanna Geiß}
\affiliation{%
  \institution{Heidelberg University}
  \streetaddress{Im Neuenheimer Feld 205}
  \city{Heidelberg} 
  \postcode{69120}
  \country{Germany} 
}
\email{geiss@informatik.uni-heidelberg.de}

\author{Michael Gertz}
\affiliation{%
  \institution{Heidelberg University}
  \streetaddress{Im Neuenheimer Feld 205}
  \city{Heidelberg}
  \postcode{69120}
  \country{Germany} 
}
\email{gertz@informatik.uni-heidelberg.de}

\renewcommand{\shortauthors}{Erich Schubert et al.}
\fi

\begin{abstract}
Many word clouds provide no semantics to the word placement,
but use a random layout optimized solely for aesthetic purposes.
We propose a novel approach to model word significance and
word affinity within a document, and in comparison to a
large background corpus. 
We demonstrate its usefulness for generating more meaningful
word clouds as a visual summary of a given document.

We then select keywords based on their significance and
construct the word cloud based on the derived affinity.
Based on a modified t-distributed stochastic neighbor embedding (t-SNE), 
we generate a semantic word placement.
For words that cooccur significantly, we include edges,
and cluster the words according to their cooccurrence.
For this we designed a scalable and memory-efficient sketch-based
approach usable on commodity hardware
to aggregate the required corpus statistics needed
for normalization, and 
for identifying keywords 
as well as significant cooccurences.
%
We empirically validate our approch using a large Wikipedia corpus.

\end{abstract}

%
%

\keywords{Word cloud, semantic similarity, keyword extraction, text summarization, visualization, sketching, hashing, data aggregation}

\makeatletter
\makeatother
\maketitle
\makeatletter
\makeatother


\section{Introduction}

\begin{figure*}[tbp]
\begin{subfigure}{.49\textwidth}\centering
  \includegraphics[width=.9\linewidth]{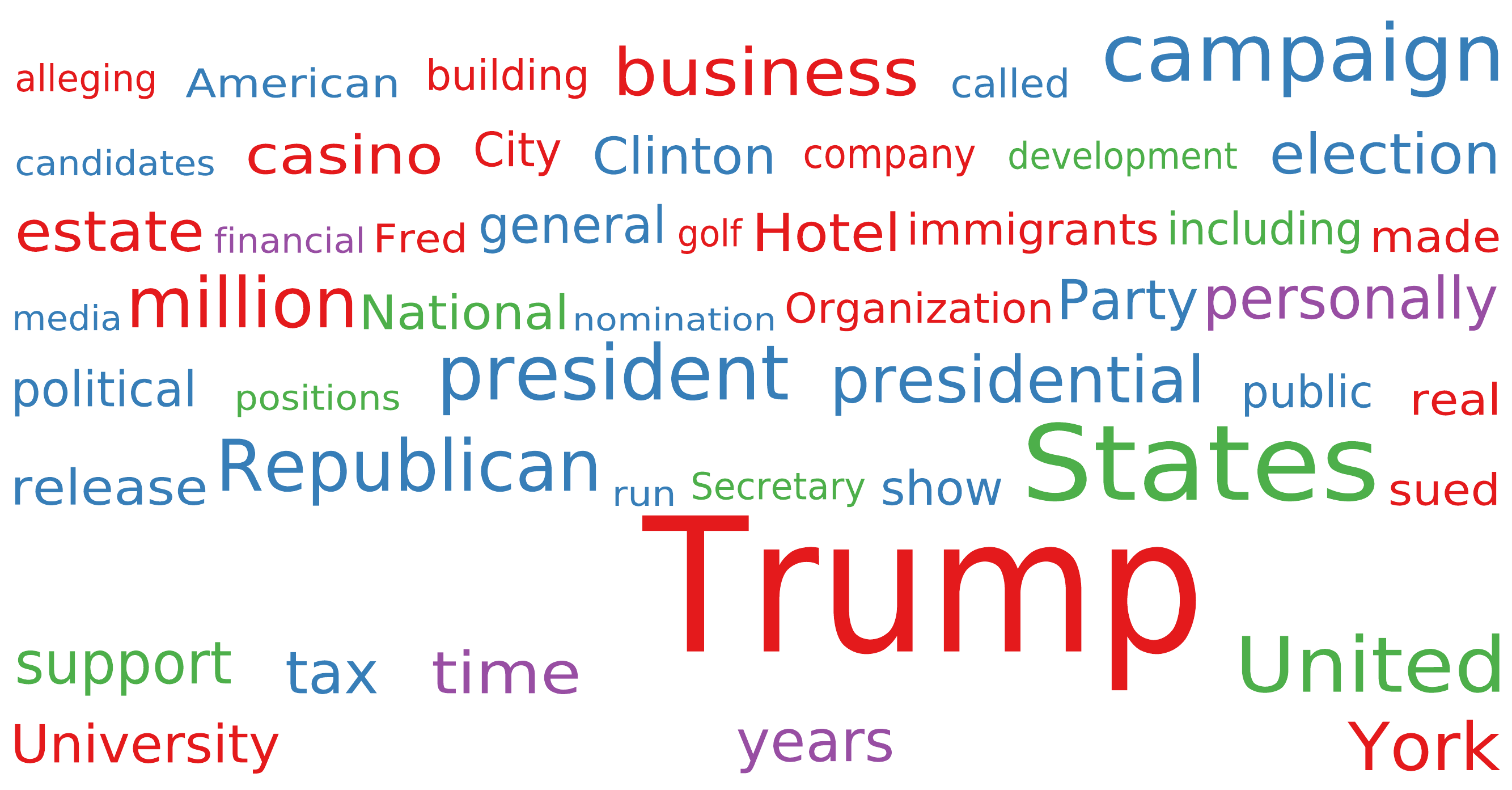}
  \caption{Alphabetic layout}
  \label{fig:trump-alpha}
\end{subfigure}
\hfill
\begin{subfigure}{.49\textwidth}\centering
  \includegraphics[width=.9\linewidth]{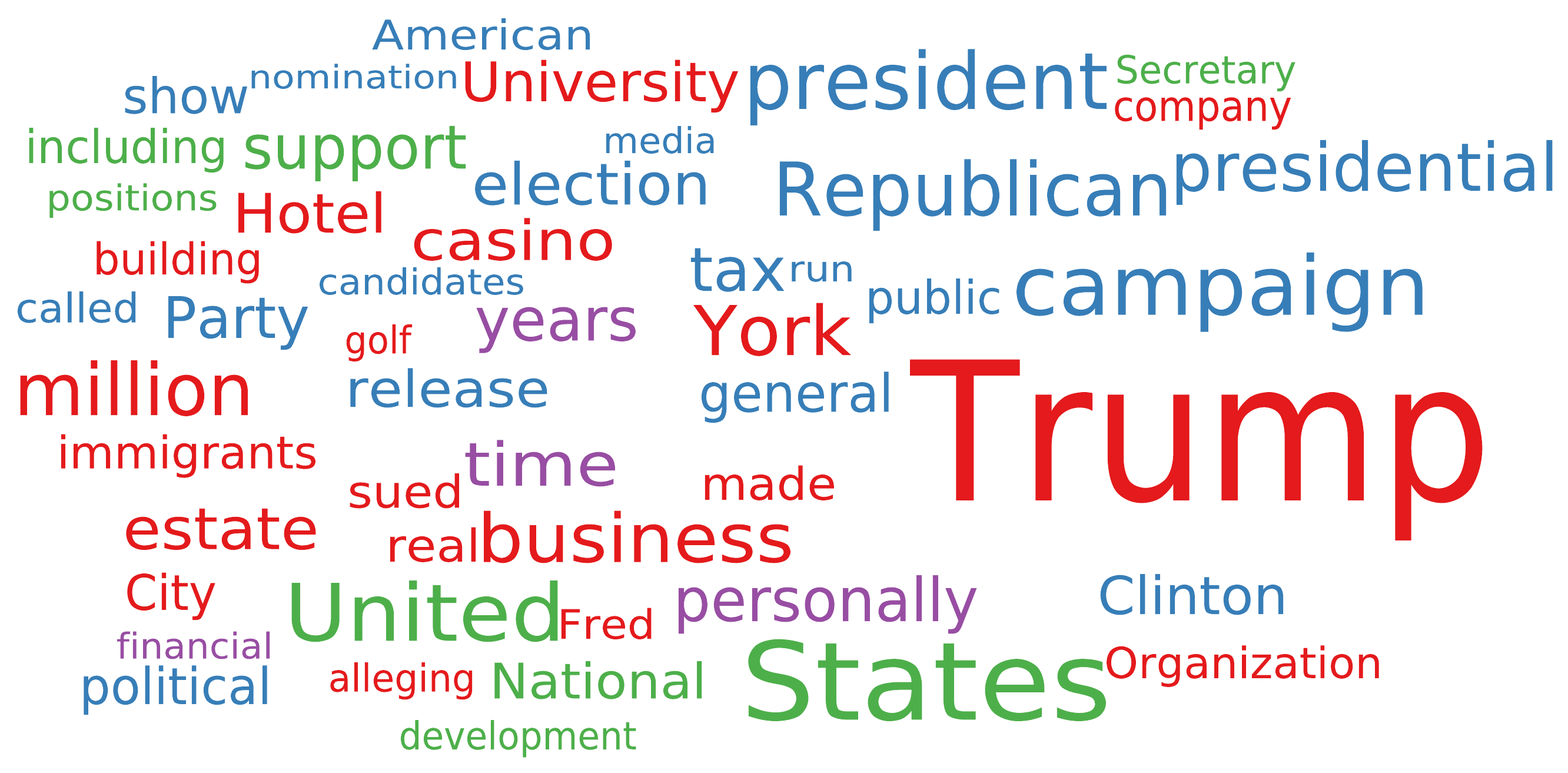}
  \caption{Random (Wordle) layout \cite{DBLP:journals/tvcg/ViegasWF09}}
  \label{fig:trump-random}
\end{subfigure}
\begin{subfigure}{.49\textwidth}\centering
  \includegraphics[width=.9\linewidth]{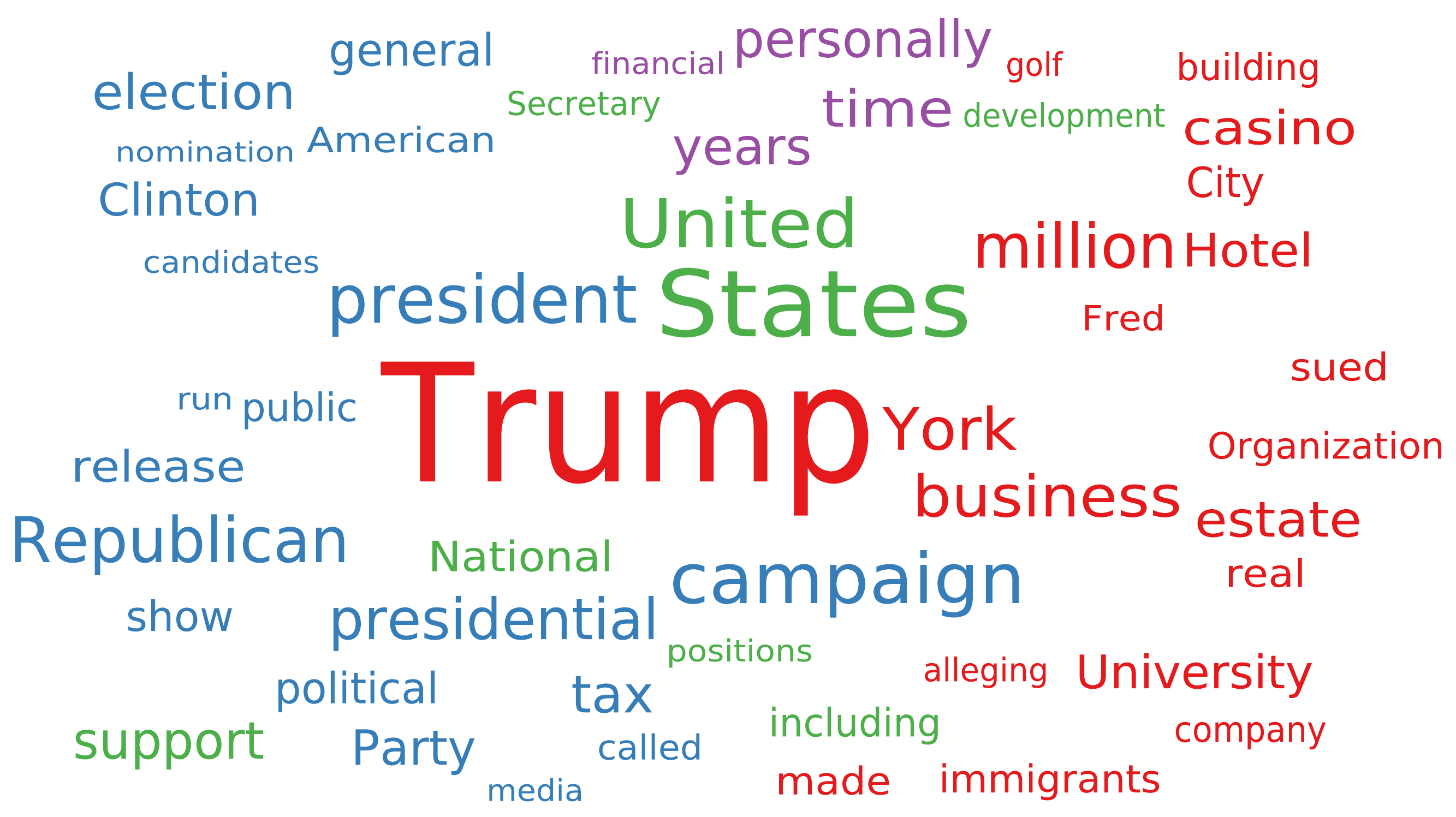}
  \caption{Force directed layout \cite{DBLP:journals/cga/CuiWLWZQ10}}
  \label{fig:trump-force}
\end{subfigure}
\hfill
\begin{subfigure}{.49\textwidth}\centering
  \includegraphics[width=.9\linewidth]{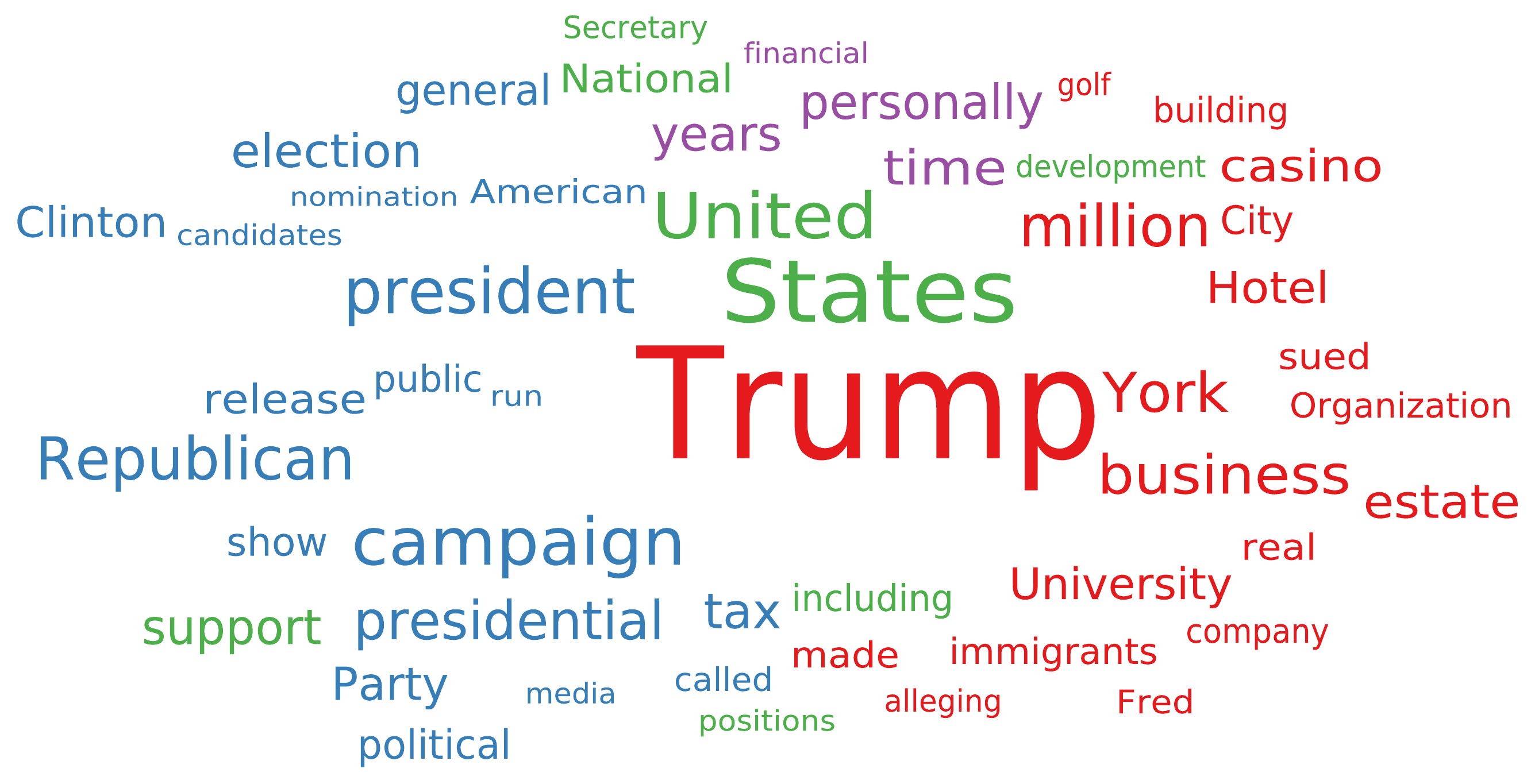}
  \caption{Seam Carving \cite{DBLP:journals/cgf/WuPWLM11}}
  \label{fig:trump-seamcarving}
\end{subfigure}
\begin{subfigure}{.49\textwidth}\centering
  \includegraphics[width=\linewidth]{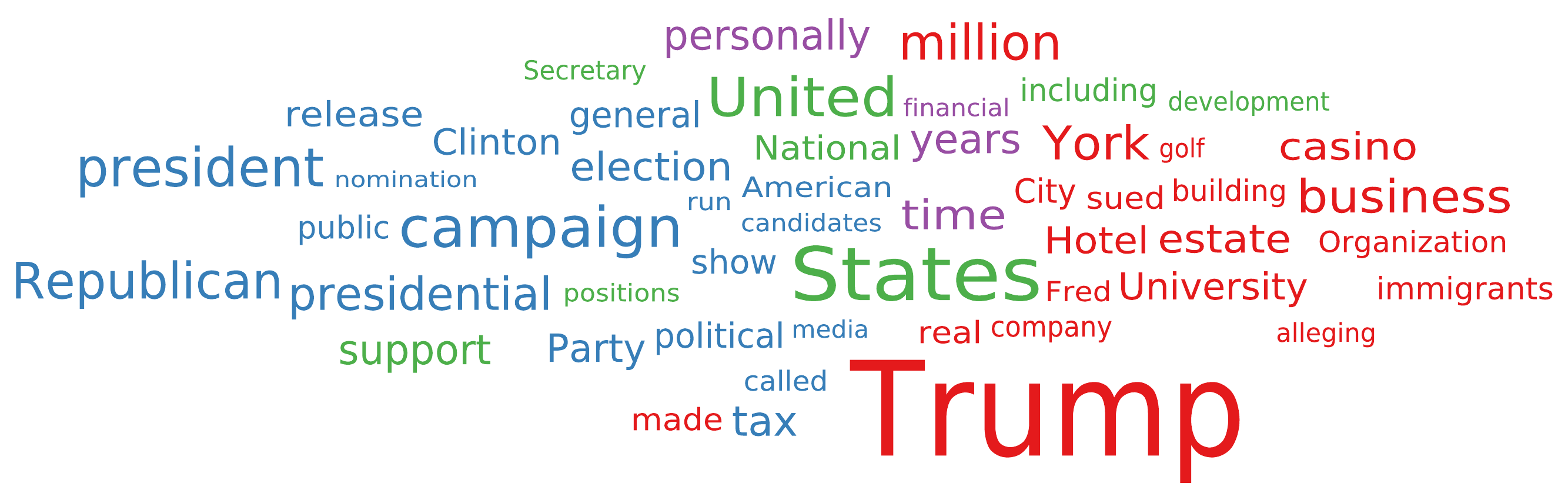}
  \caption{Inflate-push algorithm \cite{DBLP:conf/wea/BarthKP14}}
  \label{fig:trump-inflate}
\end{subfigure}
\hfill
\begin{subfigure}{.49\textwidth}\centering
  \includegraphics[width=\linewidth]{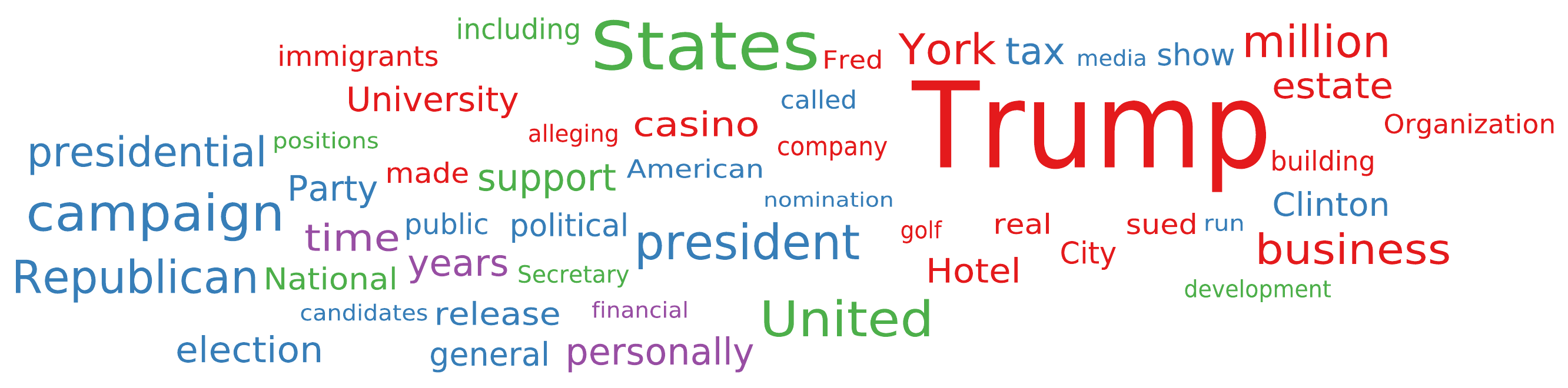}
  \caption{Star-based layout \cite{DBLP:conf/wea/BarthKP14}}
  \label{fig:trump-star}
\end{subfigure}
\begin{subfigure}{.49\textwidth}\centering
  \includegraphics[width=.9\linewidth]{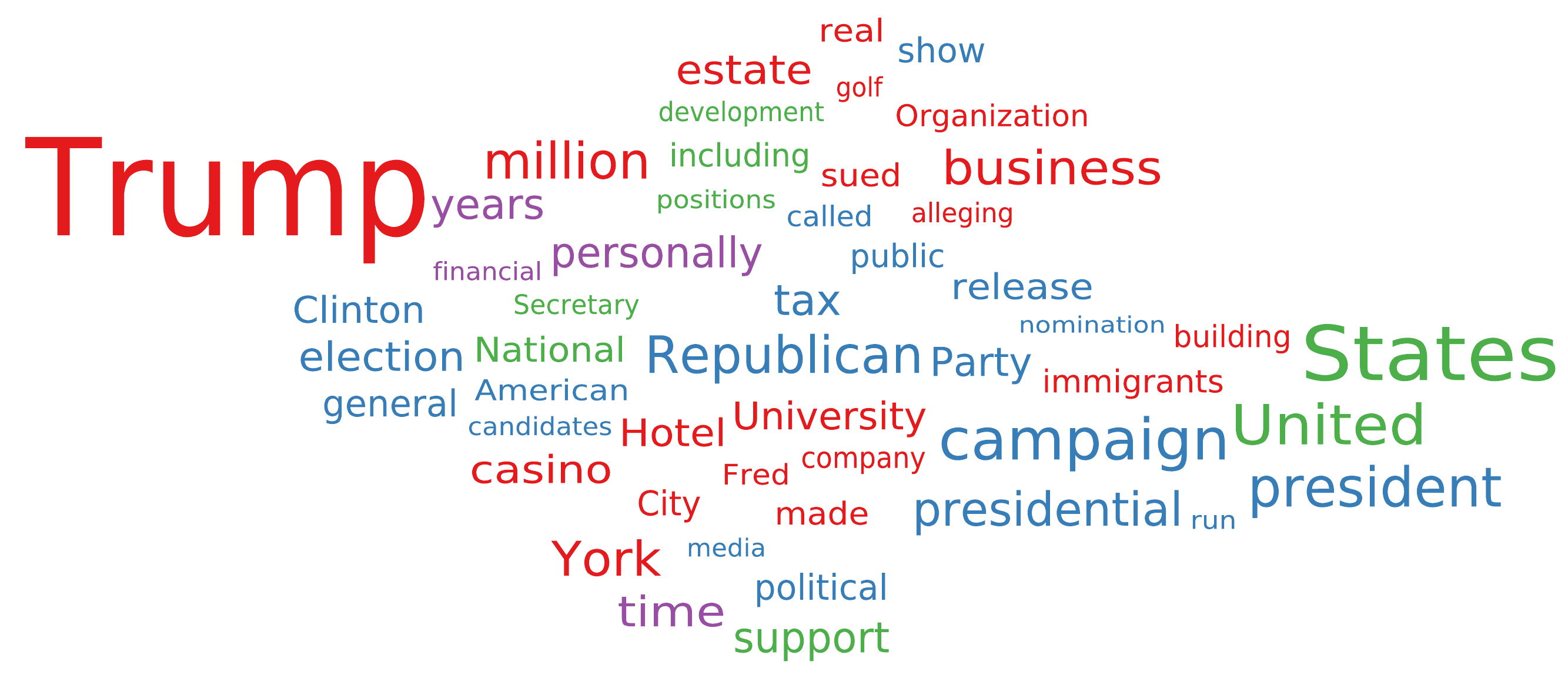}
  \caption{Cycle-cover layout \cite{DBLP:conf/wea/BarthKP14}}
  \label{fig:trump-cycle}
\end{subfigure}
\hfill
\begin{subfigure}{.49\textwidth}\centering
  \includegraphics[width=\linewidth]{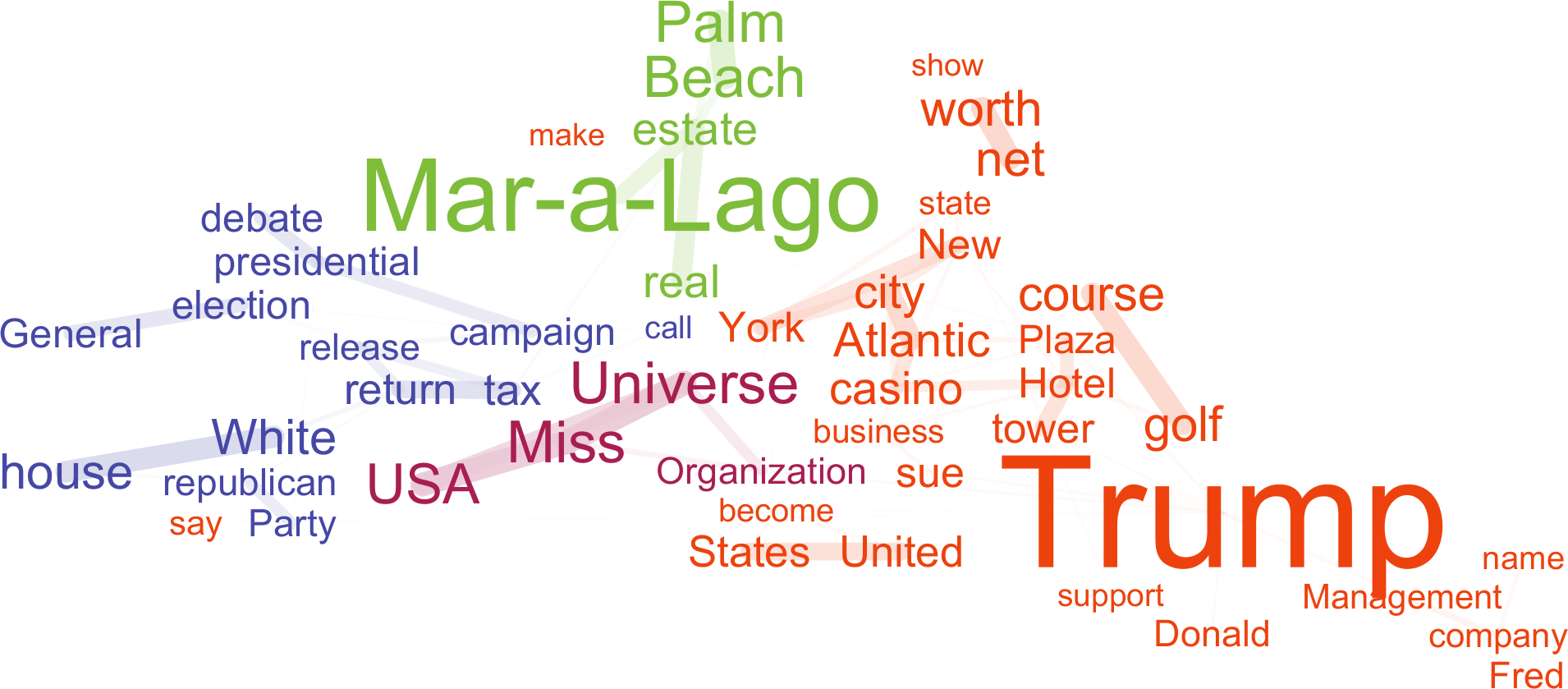}
  \caption{Semantic layout using t-SNE}
  \label{fig:trump-tsne}
\end{subfigure}
\caption{Comparison of different layout algorithms applied to the Donald Trump Wikipedia article
as of Febuary 17 2017,
generated with \url{http://wordcloud.cs.arizona.edu/} (except \reffig{fig:trump-tsne}, which is the new algorithm).
Notice that word pairs such as ``United States'' or ``presidential campaign'' are often not adjacent.
The Wordcloud tool also removes the stopword ``New'' from ``New York City''.
Our significance-base term selection method includes pairs such as ``Miss Universe'', ``Palm Beach'', and ``White House''
and emphasizes these relationships with line connections.
The force-directed algorithms usually preserve cluster structures better
than the star-based and cycle-cover algorithms.}
\label{fig:trump}
\end{figure*}

Word clouds are a popular tool for generating teaser images of textual content that serve as visual summarizations.
Originally applied to tags of image collections and blog contents, they
visually present the frequency distribution of words within a data set. By design, they are not intended to be read sequentially,
but rather only visually scanned.
Larger and more central words attract more attention.
After having been very popular because of their novelty in the first decade of this century,
the popularity of word clouds has since declined due to overuse, and because they usually only encode frequency information
but not the relationship of words. In experiments, this was not found to be always useful~\cite{DBLP:journals/jis/SinclairC08}.

The first generation tag clouds were sorted alphabetically (c.f.\ \reffig{fig:trump-alpha}),
and only vary the font size based on frequency.
The alphabetic ordering helps navigation purposes,
because the user can easily locate a known tag or verify its absence~\cite{DBLP:conf/interact/LohmannZT09}.
Many second generation variations (as shown in \reffig{fig:trump-random})
place words based on an inside-out spiral pattern in decreasing frequency,
pushing words out as far as necessary to avoid overlap.
To reduce unused space, words may also be rotated \cite{DBLP:journals/tvcg/KohLKS10}
or omitted (filling the gaps with less frequent ones instead).
Many tools also allow filling custom shapes such as logos with a word cloud,
demonstrating the use as computer art rather than a visualization technique.
Third generation ``semantic'' word clouds try to place related words 
close to each other.
This adds the 
challenges of optimizing such a semantic placement as well as measuring the similarity of words. 

Our new method improves over such third generation methods by using a
large background corpus for normalization of word frequencies,
a \emph{novel distributional probability} to measure similarity and choose terms,
and an improved layout algorithm using t-SNE.
A key contribution of our approach is the use of a sketch data structure to efficiently
store a summary of the corpus, allowing this method to be used efficiently on a computer
with limited resources, rather than requiring a complete index of the background corpus.

We first discuss related work (\refsec{sec:related}), then introduce our novel significance score
in \refsec{sec:scoring}. We show how the background data can be efficiently managed with approximate
database summarization techniques.
Because we can score both words and word interactions, this affinity can be plugged into t-SNE
and used for visualization.
We perform an empirical evaluation of our contributions in \refsec{sec:eval},
and conclude with final remarks in \refsec{sec:conclusions}.

\begin{table*}[t]{\setlength{\tabcolsep}{2pt}\small
\begin{tabular}{>{\raggedright}p{1.8cm}|p{1.3cm}|p{1.6cm}|p{2.1cm}|p{1.4cm}|>{\raggedright}p{2.cm}|p{3.7cm}}
Method & Input & Term Sel. & Features & Similarity & Initial Layout &
Postprocessing
\\
\hline
Cui et al. \cite{DBLP:journals/cga/CuiWLWZQ10}
& Document
& tf
& Cooccurrences
& Cosine
& MDS, Delaunay
& force-directed
\\
Wu et al. \cite{DBLP:journals/cgf/WuPWLM11}
& Document
& LexRank
& Cooccurrences
& Cosine
& Cui et al. \cite{DBLP:journals/cga/CuiWLWZQ10}
& + seam carving
\\
Barth et al.~\cite{DBLP:conf/wea/BarthKP14}
& Document
& tf, tf-icf,\newline LexRank
& Cooccurrences
& Cosine or\newline Jaccard
& MDS
& Inflate-and-push or star forest or cycle-covers, then force directed
\\
\hline
Adä et al.~\cite{DBLP:conf/smc/AdaTB10}
& Corpus
& tf-idf
& Document vector
& Cosine
& MDS (modified)
& force-directed with force transfer
\\
Wang et al.~\cite{DBLP:conf/graphicsinterface/WangZGNR14}
& Corpus
& tf-idf
& NLP dependency graph
& Edge tf-idf
& LinLogLayout (energy-based force-directed)
& spiral adjustment
\\
Le et al.~\cite{DBLP:conf/ijcai/LeL16}
& Corpus
& tf
& Cooccurrences
& -
& Latent model
& Spiral adjustment
\\
Xu et al.~\cite{DBLP:conf/apvis/XuTL16}
& Corpus
& tf-idf
& word2vec
& Cosine
& MDS
& Force-directed
\\
\hline
proposed
& Document vs. Corpus
& Relative (co-) occurrence
& Relative\newline cooccurrences
& Probabilistic similarity
& t-SNE (gradient descent)
& Gravity compression
\end{tabular}
\caption{Third-generation Word-Cloud approaches compared to the proposed method}
\label{tab:thirdgen}
}
\vspace{-2ex}
\end{table*}

\pagebreak
\section{Related Work}\label{sec:related}

We give a short overview of word clouds,
layout algorithms commonly used for word clouds,
and the t-stochastic neighbor embedding (t-SNE) method that we employ for our visualization.

\subsection{Word Clouds}\label{subsec:wordclouds}

Word clouds (originally ``tag clouds'') are a visual quantitative summary of the frequency
of words in a corpus. In the basic form, the most frequent words are presented
as a weighted list with the font size scaled to emphasize more frequent terms.

An early example of a semantic word cloud can be found in \cite{MilgramJ76}, who
drew a map of Paris placing the names of landmarks with their font size scaled to reflect the frequency in hand-drawn maps.
Word clouds became popular
when websites such as Flickr used them to show tag popularity,
and tools such as Wordle~\cite{DBLP:journals/tvcg/ViegasWF09} made it easy for
everybody to generate a word cloud from his own text. This second generation of word clouds places
words randomly to fill a given area or polygon \cite{DBLP:conf/iv/SeifertKKGG08},
and focuses on generating ``art''. 

This early use of word clouds for website navigation is not without criticism:
Sinclair et al.~\cite{DBLP:journals/jis/SinclairC08} found that traditional search-based interfaces
were more useful for searching specific information, as the word cloud may not contain the desired terms,
and may only allow access to parts of the data, but admit that it can provide a good visual summary of the database.
Heimerl et al.~\cite{DBLP:conf/hicss/HeimerlLLE14} note that word clouds only represent a
``purely statistical summary of isolated words without taking linguistic knowledge
[\ldots] into account''~\cite{DBLP:journals/tvcg/ViegasWHKM07}.
Several user studies found that alphabetic ordering works well for locating tags,
while explorative tasks are usually better supported by semantic
layouts~\cite{DBLP:conf/interact/LohmannZT09,DBLP:conf/interact/DeutschST09,DBLP:conf/chi/SchrammelLT09}.
Force-directed graphs were introduced by Tutte~\cite{Tutte63}, popularized by
Fruchterman et al.~\cite{DBLP:journals/spe/FruchtermanR91},
and applied to layouting tag clouds, e.g., in~\cite{DBLP:journals/ijcicg/ChenSBT10}.

Third-generation word cloud approaches mostly use term frequency for selecting terms,
then usually use cosine to measure the similarity of terms based on cooccurrences, MDS to produce
the initial layout, and a force-directed graph to optimize the final layout.
\reftab{tab:thirdgen} gives an overview of third-generation techniques and their differences.
Overviews of related work can be found, e.g., in \cite{DBLP:conf/wea/BarthKP14,DBLP:conf/hicss/HeimerlLLE14}.

If only a single document is available, there is usually too little data to use for choosing words, 
except taking the most frequent words (excluding stopwords). Some implementations can use
the inverse in-corpus frequency (often with respect to the Brown corpus) for weighting.
Only methods that visualize an entire corpus can use the \textrm{tf-idf} combination that is popular in
text search. Cooccurrence vectors are then either computed at the sentence level (single document)
or document level (corpus visualization).

\subsection{Layout Algorithms}\label{subsec:layout}

The initial layout is usually obtained by multidimensional scaling (MDS), but this suffers from
the problem that the observed similarity values are too extreme.
On one hand, words will be mapped to almost the same coordinates, thus causing overlap.
On the other hand, because MDS tries to best represent the far distances,
MDS tends to cluster the unique words in different corners of the data space~\cite{DBLP:conf/smc/AdaTB10},
as this best represents dissimilarity in the data set.
Because of this, all authors chose to perform some kind of postprocessing to reduce the amount
of unused space, and to avoid word overlaps. Common solutions here include the use of force-based
techniques that repel overlapping words, and close gaps by attractive forces.

Some methods use word embedding such as word2vec and latent variable models~\cite{DBLP:conf/apvis/XuTL16,DBLP:conf/ijcai/LeL16}.
These models require a substantial amount of data, and therefore need to be computed on the
entire corpus rather than a single document. The basic layout will then be the same for all
documents, and only vary by term selection and postprocessing.
We want a layout that is focused on the significant interactions within a single document,
and therefore do not consider global word embeddings: term interactions are of the highest
interest if they differ from the corpus-wide behavior. Because of this, we do not consider
global word embeddings such as word2vec~\cite{DBLP:journals/corr/abs-1301-3781} and
GloVe~\cite{DBLP:conf/emnlp/PenningtonSM14} to be viable alternatives here.

\subsection{t-Stochastic Neighbor Embedding}\label{subsec:tsne}

Stochastic neighbor embedding (SNE)~\cite{DBLP:conf/nips/HintonR02} and t-distributed
stochastic neighbor embedding (t-SNE)~\cite{journals/jmlr/MaatenH08}
are projection techniques designed for visualizing high-\dimensional{} data in a low-\dimensional{} space (typically 2 or 3 dimensions).
These methods originate from computer vision and deep learning research, where they are used to visualize large image collections.
In contrast to techniques such as principal
component analysis (PCA) and multidimensional scaling (MDS), which try to maximize the spread of dissimilar
objects, SNE focuses on placing similar objects close to each other, i.e., it preserves locality rather than
distance or density.

The key idea of these methods is to model the high-\dimensional{} 
input data
with an affinity probability distribution, and use gradient descent to optimize the
low-\dimensional{} projection to exhibit similar affinities.
Because the affinity has more weight on nearby points, we obtain a non-linear projection that
preserves local
neighborhoods, while it can move away points rather independently of each other. 
In SNE, Gaussian kernels are used in the projected space, whereas t-SNE uses a Student-t distribution.
This distribution is well suited for the optimization procedure because it is computationally inexpensive,
heavier-tailed, and has a well-formed gradient.
The heavier tail of t-SNE is beneficial, because it increases the tendency of the projection to
separate unrelated points in the projected space.

In the input domain, SNE and t-SNE both use a Gaussian kernel for the input distribution.
Given a point $i$ with coordinates $\V{x}_i$, the conditional probability density $\pjGi$ of any neighbor point $j$ is computed as
\begin{equation}
\pjGi =
\tfrac{ \phantom{\sum_{k\neq i}}%
\exp(-\norm{\V{x}_i-\V{x}_j}^2 / 2\sigma^2_i)}%
{\sum_{k\neq i}\exp(-\norm{\V{x}_i-\V{x}_k}^2 / 2\sigma^2_i)}
\label{eqn:sne-p}
\end{equation}
where $\norm{\V{x}_i-\V{x}_j}$ is the Euclidean distance,
and the kernel bandwidth $\sigma_i$ is optimized for every point
to have the desired perplexity (an input parameter roughly
corresponding to the number of neighbors to preserve).
The symmetric affinity probability $\pij$ is then obtained as
the average of the conditional probabilities $\pij = \frac{1}{2}(\piGj+\pjGi)$ and
is normalized such that the total sum is $\sum_{i\neq j} \pij = 1$.
SNE uses a similar distribution (but with constant $\sigma$)
in the projected space of vectors~$\V{y}_i$, whereas t-SNE uses the Student-t distribution instead:
\begin{equation}
\qij =
\tfrac{ \phantom{\sum_{k\neq l}}%
(1 + \norm{\V{y}_i-\V{y}_j}^2)^{-1}}%
{\sum_{k\neq l}(1 + \norm{\V{y}_k-\V{y}_l}^2)^{-1}}
\label{eqn:tsne-q}
\end{equation}
The denominator normalizes the sum to a total of $\sum_{i\neq j} \qij = 1$.
The mismatch between the two distributions can now be measured using the Kullback-Leibler divergence
\begin{equation}
\operatorname{KL}(P\mid\mid Q) :=
\sum\nolimits_i\sum\nolimits_j \pij \log \tfrac{\pij}{\qij}
\end{equation}
To minimize the mismatch of the two distributions,
we can use the vector gradient $\tfrac{\delta C}{\delta \V{y_i}}$ (for Student-t / t-SNE, 
c.f.\ \cite{journals/jmlr/MaatenH08}):
\begin{equation}
\tfrac{\delta C}{\delta \V{y_i}} :=
4 \sum\nolimits_j (\pij-\qij)\,\qij\,Z\,(\V{y}_i-\V{y}_j)
\label{eqn:tsne-grad}
\end{equation}
where $Z=\sum_{k\neq l}(1+\norm{\V{y}_k-\V{y}_l}^2)^{-1}$
(c.f.~\cite{DBLP:journals/jmlr/Maaten14}).
Starting with an initial random solution, the solution is then iteratively optimized using
gradient descent with learning rate $\eta$ and momentum $\alpha$ (c.f.~\cite{journals/jmlr/MaatenH08}):
\begin{equation}
Y_{t+1} \leftarrow
Y_t - \eta \tfrac{\delta C}{\delta Y}
+ \alpha \left(Y_t - Y_{t-1}\right)
\end{equation}

There are interesting similarities between this optimization and force-directed graph drawing algorithms.
For example, a similar momentum term is often used; and we can 
interpret
the gradient \refeqn{eqn:tsne-grad} as attractive forces (coming from the similarity in the input space~$\pij$)
and repulsive forces (from the affinity in the projected space,~$-\qij$).
The affinity in the output space $\qij$ also serves as a weight applied to the forces,
i.e., nearby neighbors will exercise more force than far away neighbors.
Where traditional force-directed graphs are based on the physical intuition of springs that
try to achieve the preferred distance of any two words
(with an attractive force if the distance is too large, and a repulsive force if the distance is too small),
SNE is based on the stochastic concept of probability distributions,
and tries to make the input and the projection affinity distributions similar.
In particular, if the input and output affinities agree ($\pij=\qij$), then we get zero force contribution,
but the forces also generally drop with the distance due to the $\qij$ factor.
Note that this factor was not heuristically added, but arises from the derivation of the Student-$t$ distribution.

\section{Significance Scoring}\label{sec:scoring}

The original t-SNE method was designed for high-\dimensional{} point data such as images.
In order to apply them for visualizing the relationships of words,
we have to replace \refeqn{eqn:sne-p} with a
probability measure that captures the desired affinity of words.

Rather than using cosine as a distance measure, our idea is to directly work with a notion of
affinity, based on the \emph{significant} cooccurrence of words.
In Natural Language Processing (NLP), the notion of affinity between words is well established,
and has long been studied and employed in text and word similarity analysis~\cite{DBLP:journals/csl/DaganMM95}.
Our work is focused around a novel significance measure, and the efficient computation
of this measure using a hashing-based summarization of the background corpus. 
Wikipedia, in particular, has been demonstrated to be an invaluable training resource for
such contextual similarities of words in actively used language~\cite{DBLP:journals/jair/PonzettoS07}
as well as implicit word and entity relationship networks in information retrieval \cite{DBLP:conf/asunam/GeissSG15,DBLP:conf/sigir/SpitzG16}.

We first define a novel significance measure for word (co-) occurrences,
with a careful Laplace style correction, which can be transformed into a
probability for the subsequent steps. Secondly, we select the most important
words based on this measure. Third, we use these probabilities to project
the data with t-SNE, which needs some subtle modification. We then introduce
the postprocessing step to reduce the amount of white space in the plot,
and use the cooccurrence probabilities to cluster the words; and give some
technical details on the rendering method used.
In \refsec{subsec:approx} we introduce a sketching technique to improve the
scalability of our method.

\subsection{Significance of Word Cooccurrences}\label{subsec:significance}

To quantify the cooccurrence of words, we process the input text using
CoreNLP~\cite{DBLP:conf/acl/ManningSBFBM14}. We split the document into sentences,
each sentence into tokens $(t_1,\ldots,t_l)$, and lemmatize these tokens.
We only consider verbs, nouns and adjectives 
for inclusion.
Many common stop words are either removed by this part-of-speech filter,
or they will usually not be significant (and if they are significant, then we want to include them).
We use only three stop words: ``be'', ``do'', and ``have'' that are common auxiliary verbs.
Because we use lemmatization, we sometimes include similar words, such as ``president''
and ``presidential'', and ``science'' and ``scientific''. If this is not desired,
e.g., stemming could be used instead of lemmatization, or even a combination of both.

To emphasize nearby words,
we use a Gaussian weight of $w_d = \exp(-d^2/2\sigma^2)$ for words at a distance of $d$
within the document~$D$, or a collection of documents~$C$.
In our experiments, we use $\sigma=4$, which gives the weights
$w_1\ldots w_8 \approx 0.97, 0.88, 0.75, 0.61, 0.46, 0.32, 0.22, 0.14$;
a~similar weighting scheme on sentences
rather than words was used, e.g., in~\cite{DBLP:conf/sigir/SpitzG16}.
The Gaussian scaling term $1/\sqrt{2\sigma^2\pi}$ can be omitted as it will cancel out after normalization.
We aggregate the weighted cooccurrences over all sentences for any two words $a$~and~$b$:
\begin{equation}
c_D^\prime(a, b) := \sum\nolimits_{S=(t_1,\ldots,t_n)} \sum\nolimits_{\substack{i<j\\t_i=a\enskip t_j=b}} \exp(-|i-j|^2/2\sigma^2)
\end{equation}
Finally, we normalize such that the total sum is $\sum_{i<j} c_D(i,j) = 1$:
\begin{equation}
c_D(a,b) = c_D^\prime(a,b)/\Sigma_D \qquad \text{ with } \Sigma_D:=\sum\nolimits_{i<j}c_D^\prime(i,j)
\end{equation}
It is sufficient to compute this for all $a<b$, because $c_D(a,b)=c_D(b,a)$.
$c_C$ is defined the same with respect to $C$.

The resulting empirical cooccurrence probabilities do not yet capture how unusual this
occurrence is. For example, the two words ``this'' and ``is'' are very likely to
cooccur in any English text and would thus have a large $c_D$ value,
but are of little interest for visualization.
Therefore, we now look at the likelihood ratio of the word pair $(a,b)$
occurring in document~$D$ as opposed to the background corpus~$C$.
Because our document is usually much smaller than our background corpus, we
need to perform a Laplacian-style correction for unseen data.
For this, we use a small weight of $\beta_D=\tfrac{1}{2}/\Sigma_D$ to account
for one word pair being included or missing just by chance (the factor $\tfrac{1}{2}$ is to account for
the Gaussian weighting applied to pair occurrences).
We reduce the weight obtained from the document by $\beta_D$,
and also increase the weight obtained from the corpus by a corpus-dependent constant $\beta_C$.
Without the $\beta$ terms, any word that does not exist in the corpus would achieve an infinite score.
We then obtain the odds
\begin{equation}
r_{ab} := \max\left(
\tfrac{c_D(a,b) - \beta_D}{c_C(a,b) + \beta_C}, 0\right) \cdot \textrm{prior}
\label{eqn:ratio}
\end{equation}
In particular, the $\beta$ adjustment accounts for words that have not been
seen before, such as spelling errors.
A similar term was previously used for the same purpose in change detection
on textual data streams by \cite{DBLP:conf/kdd/SchubertWK14,DBLP:conf/ssdbm/SchubertWK16}.
As $\textrm{prior}$ ratio we simply use
\begin{equation*}
\textrm{prior} = k/|W|
\end{equation*}
where $k$ is the number of words to chose,
and $|W|$ is the number of unique words in the document. This parameter is a constant for the
document, and thus does not affect the ranking of words, only the absolute score values.
An intuitive interpretation of this ratio is to estimate the odds of a set of words being better
suited for describing the current document rather than another document in the corpus.
To transform the ratio into a probability, we use the usual odds to probability conversion:
\begin{equation}
p_{ab} :=
\tfrac{r_{ab}}{r_{ab}+1}
\label{eqn:prob}
\end{equation} 
This transformation is monotone and is only required when we need a probability value $p_{ab}$
(e.g., for applying the SNE algorithm),
while for selecting words (and font sizing) we use the ratio values~$r_{ab}$ as explained in the next section.

In contrast to existing measures such as pointwise mutual information (PPMI)~\cite{DBLP:journals/coling/ChurchH90},
our method does not assume independence, but rather we rely on the empirical cooccurrence frequency from a
large document corpus as normalizer. In \refsec{subsec:approx} we will introduce a compact summary to efficiently
store and retrieve these frequencies, which makes feasible to use this method on a desktop computer.
Because of this, the associations discovered by our approach are more
relevant to the document itself.

\subsection{Selecting Keywords for Inclusion}
\label{subsec:words}

To select word pairs for inclusion, we score them by \refeqn{eqn:ratio}
(or equivalently their probability, \refeqn{eqn:prob}).
We may also want to select single words that occur unusually often using a single-word version of the same equations:
\begin{align}
c^\prime_D(a) :=& 
\sum\nolimits_{S=(t_1,\ldots,t_n)} \sum\nolimits_{\substack{i}} \mathbbm{1}_{t_i=a}
\\
c_D(a) :=& c^\prime_D(a) / \Sigma_w \qquad\qquad\quad\enskip\text{ with }\Sigma_w := \sum\nolimits_i c^\prime_D(i)
\label{eqn:weight-word}
\\
r_a :=& \max\left( \tfrac{c_D(a) - \beta^\prime_D}{c_C(a) + \beta_C}, 0\right)
\quad\quad\!\text{with }
\beta^\prime_D := 1/\Sigma_w 
\label{eqn:score-word}
\end{align}
where $c_D(a)$ and $c_C(a)$ are the empirical probability of word $a$ occurring in the document $D$
and the corpus $C$, respectively. Here, $\beta^\prime_D$ is simply the inverse of the total document weight, because we
did not apply weights to the individual word occurrences.

We can then select the desired number of words (either top-$k$, and/or with a threshold)
based on the decreasing maximum likelihood of the word itself
and of all pairs it participates in:
\begin{equation}
s_a := \max\left\{ r_a, \enskip \max\nolimits_i r_{ai} \right\}
\label{eqn:score-max}
\end{equation}

\subsection{Modifications to t-SNE for Improved Layout}

\begin{figure}[tb]\centering
\begin{subfigure}[b]{.49\textwidth}\centering
  \includegraphics[width=\linewidth]{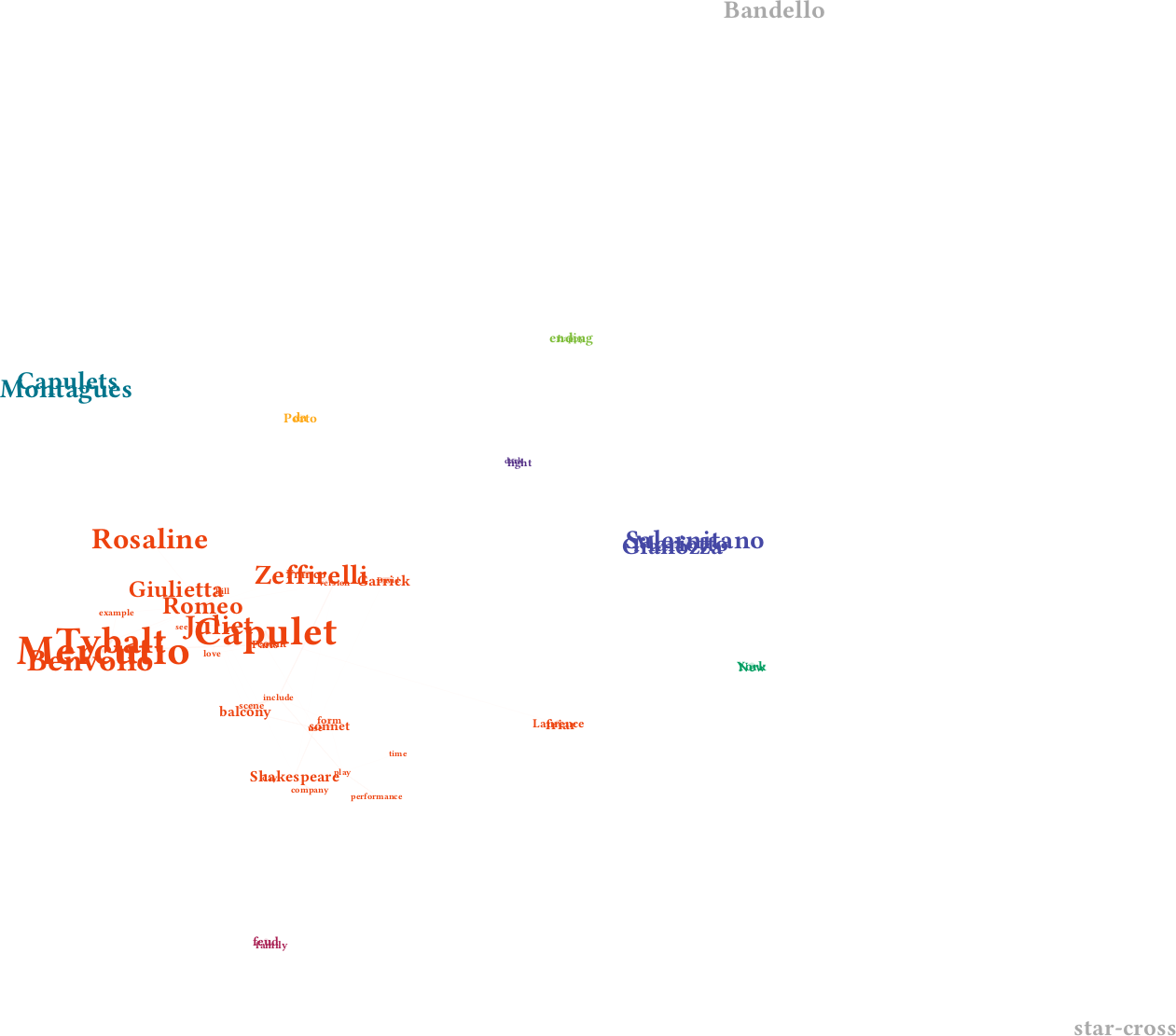}
  \caption{t-SNE word cloud without compression.
  There is a lot of white space and too small font sizes although we allow words to overlap.
  }
  \label{fig:optimization-nocompress}
\end{subfigure}
\begin{subfigure}[b]{.49\textwidth}\centering
  \includegraphics[width=\linewidth]{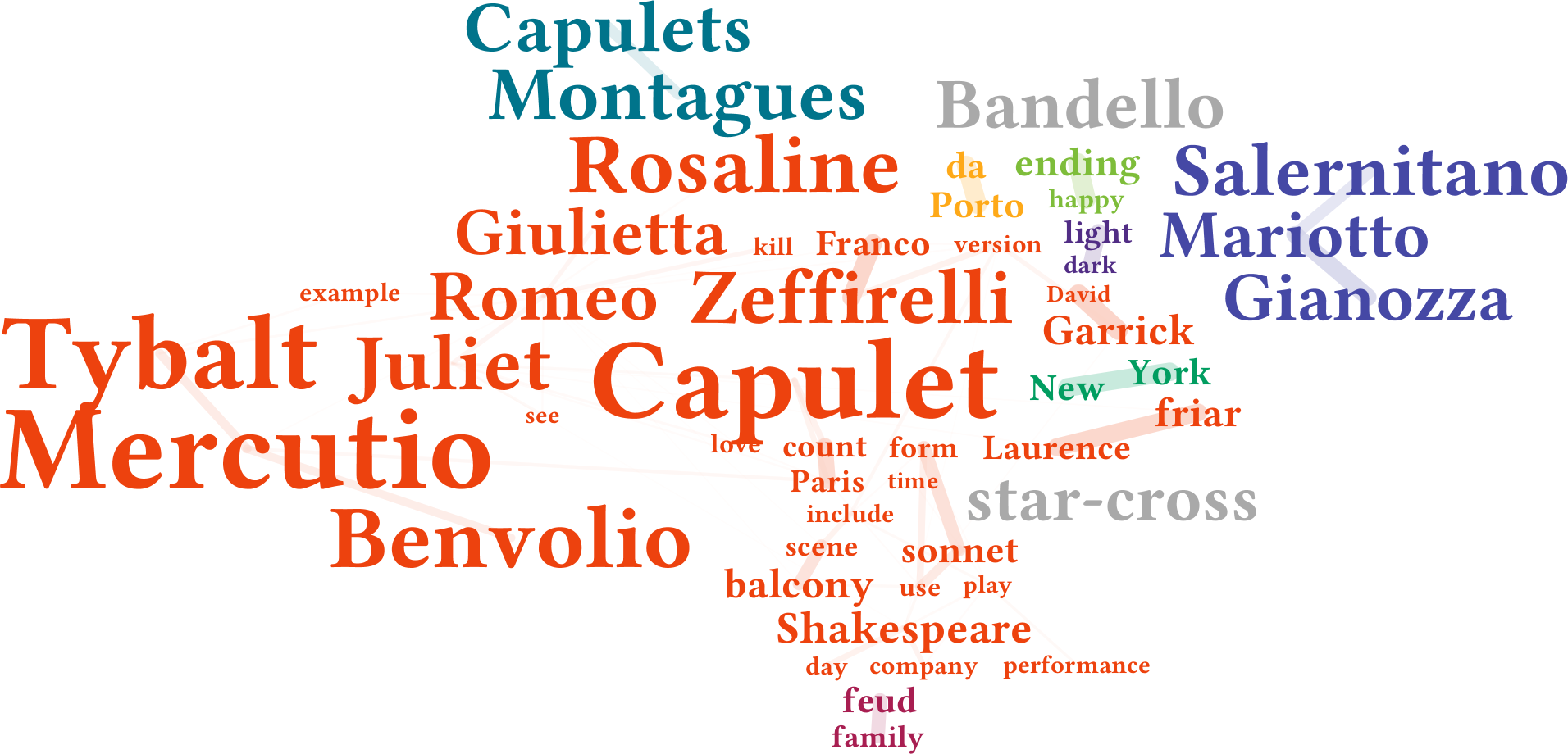}
  \caption{t-SNE word cloud with gravity compression.\\
  \phantom{ty}}
  \label{fig:optimization-full}
\end{subfigure}
\caption{Effect of layout optimization contributions on the visualization of the ``Romeo and Juliet'' Wikipedia article.
}
\label{fig:optimization}
\end{figure}

Simply applying the t-SNE projection to the probabilities $\pij$ from \refeqn{eqn:prob} will often yield a
layout that diverges too much. This is caused by many tiny probabilities (often even zero)
that create disconnected groups of words.
The optimization procedure pushes these groups further apart. In the most extreme case, we have
multiple words that occur frequently in the text, but never in the same sentence, and thus have $\pij=0$;
but this problem can also arise due to many small similarities. \reffig{fig:optimization-nocompress} shows
the result of applying t-SNE onto the unmodified $\pij$ for the ``Romeo and Juliet'' Wikipedia article.
%
%
The longer we iterate t-SNE, the more the individual components are pushed apart, and we have to use tiny font sizes,
or accept overlapping text. We originally countered this with a normalization approach that would use a general
background affinity to pull together everything to some extend, but this became eventually obsolete
when we added the gravity-based optimization introduced in \refsec{subsec:gravity}.

In general, t-SNE spreads out the words too much rather than having connected words touch each other.
This is in fact a desired behavior, as it reduces the crowding problem.
In prior work with MDS (e.g., \cite{DBLP:conf/smc/AdaTB10})
a problem was that words tend to be placed exactly on top of each other, which t-SNE avoids.
Because of this, the unmodified t-SNE projection always has a lot of white space, and very small fonts if we do not allow them to overlap
(in \reffig{fig:optimization-nocompress}, we already allow some overlap to be able to use larger fonts).
We also initially experimented with using bounding box based distances in t-SNE, and weighted gradients in order to incorporate
word sizes into the optimization procedure, but this turned out to be a dead end,
as t-SNE will then simply produce white space in-between the bounding boxes for the same reasons as above.
All we kept from these experiments is the use of a streched coordinate system using the golden ratio
$\phi=\frac{1+\sqrt{5}}{2}$ to obtain a more pleasant layout.

\subsection{Gravity-based Optimization}\label{subsec:gravity}

Because we could not optimize the layout with bounding boxes directly,
we follow the best practise and produce an initial layout of points,
then postprocess it with a 
gravity-based compression.
Alternatively, a force-directed graph, force-transfer~\cite{DBLP:conf/smc/AdaTB10}
or the seam-carving algorithm \cite{DBLP:journals/cgf/WuPWLM11} could be used instead.
Our gravity-based compression is an iterative optimization, beginning with the initial
layout obtained from t-SNE. In each iteration, the shared scale of all words is increased
as far as possible without causing word overlap.
We then use each word in turn as center of gravity,
sort all words by distance to this center, and move them slightly towards the center of gravity unless this would
cause words to overlap. Once we cannot move words any closer, the algorithm terminates.
This approach can be seen as the opposite of the inflate-and-push algorithm~\cite{DBLP:conf/wea/BarthKP14}:
our words are too much spread out by t-SNE, and we need to pull them together.
By using different centers of
gravity and only moving words a small amount, we preserve the word relationships reasonably well, although some
distortion may occur. For example, in the visualization of the ``Romeo and Juliet'' article in \reffig{fig:romeo-wikipedia}, the words ``bite thumb''
are separated by ``Juliet'' because of the gravity optimization ignoring the word relationships.

\subsection{Clustering}

A separate way of displaying word affinities is by using colorization and clustering.
We can also use the $\pij$ values to cluster the data set. For this we experimented
with hierarchical agglomerative clustering (HAC).
We tried several clustering algorithms from ELKI~\cite{DBLP:journals/pvldb/SchubertKEZSZ15},
including the promising Mini-Max clustering technique~\cite{Bien11},
because it was supposed to provide central prototype words. Unfortunately, the natural idea of having
prototypes for clusters does not appear to work well with the $\pij$,
likely because our similarity does not adhere to the triangle inequality
(a word $a$ can be frequently cooccurring with
two words $b$ and $c$ even when these never appear together).
The results with group average linkage (average pairwise affinity, UPGMA,~\cite{Sokal58})
were more convincing. This clustering algorithm computes a pairwise affinity matrix,
considers each word to be its own cluster, and then iteratively merges
the two most similar clusters.
When merging clusters, the new affinity is computed as the average of the
pairwise affinities, i.e.
\begin{equation}
s(A,B) :=
\tfrac{1}{|A|\cdot|B|} \sum\nolimits_{i\in A,\, j\in B} s(a,b)
\end{equation}
which can be computed efficiently using the Lance-Williams equations \cite{Lance67}.
The clustering algorithm runs in $\mathcal{O}(n^3)$, but for less than a thousand words the resulting
run-time is negligible because we already have the affinity matrix.

From the cluster dendrogram (the tree structure obtained from the repeated merging of clusters),
we extract up to $K=8$ clusters with at least 2~points each,
while isolated singleton points are considered ``outliers'', and represented by a gray color in the figures.

We implemented a new logic for cutting the dendrogram tree based on the idea of
undoing the last $K-1$ merges. At this point we have $K$ clusters, but some of them may contain a single point only.
If we have less than $K$ clusters of at least $2$ points,
we undo additional merges until the desired number of non-singleton clusters is found.
If this condition cannot be satisfied, we instead return the result with the largest number of such clusters,
and the least number of merges undone, i.e., if we cannot find a solution with $K=8$ clusters, we try to find
one with $K=7$ clusters and so on. This can be efficiently implemented in a single bottom-up pass over the dendrogram
by memorizing the best result seen when executing one merge at a time.

\subsection{Output generation}

Our drawing routines are basic:
Scores are linearly normalized to $[0;1]$ based on the minimum and maximum score included.
Font sizes employ the square root (because the area 
is quadratic in the font size) of the score and ensure a minimum font size of 20\%, i.e.: 
\begin{align}
\text{font-size}(i) :=&
\sqrt{\tfrac{r(i)-\min_j r(j)}{\max_j r(j)-\min_j r(j)}} \cdot 80\% + 20\%
\end{align}
We compute the bounding boxes of all words, then translate and scale the coordinates and font sizes
such that they fit the screen.
We use two layers: the bottom layer containing the connections 
between significantly cooccurring words
with reduced opacity,
and the top layer containing the words.

\subsection{Scalability Considerations}\label{subsec:approx}

The proposed scoring method requires the weighted corpus cooccurrence frequency $c_C(a,b)$ for any pair of words $(a,b)$ that may arise.
Computing this on-demand is too slow for a corpus like Wikipedia; but we would also like to avoid storing all
pairwise coocurrences: there are about 10.9~million unique verbs, nouns, and adjectives after lemmatization in Wikipedia.
Storing just
the counts of single words uses about 190~MB; counting all weighted pairwise occurrences requires several GB.
But because we
do not rely on exact values (Wikipedia changes, so any exact value would be outdated immediately), we can employ
an approximation technique based on count-min sketches~\cite{DBLP:journals/jal/CormodeM05}. These sketches work
with a finite sized hash table. We used a single table with $2^{26}$~buckets $B[i]$, and $4$~hash functions~$h_j$.
The resulting hash table (with float precision) occupies only 256~MB of disk space (the CoreNLP language model
occupies about 1 GB).
When one Wikipedia article $A\in W$ is completely processed,
the resulting counts are written to the hash table bucket $B[i]$
using the maximum of all values with the same hash code $i$:
\begin{align}
B[i] \leftarrow 
B[i] + \max\big(
& \max\nolimits_{a<b}\left\{ c_A(a, b) \mid \exists_j h_j(a,b)=i \right\},
\\
& \max\nolimits_{a\phantom{<b}} \left\{c_A(a)\phantom{,b} \mid \exists_j h_j(a)\phantom{,b}=i \right\}
\big)
\shortintertext{\noindent
To estimate the frequency of $c_C(a,b)$, we use
}
c_C(a,b) &\approx
\min \left\{B[h_j(a,b)] \mid j=1\ldots 3\right\} \big/ |W|
\\
c_C(a)\phantom{,b} &\approx
\min \left\{B[h_j(a)]\phantom{,b} \mid j=1\ldots 3\right\} \big/ |W|
\end{align}
This ``write max, read min'' strategy guarantees that we never underestimate $c_C(a,b)$.
We overestimate $c_C$ only if each of the hash functions has a collision with a more frequent word pair.
This will only be the case for rare terms, for which we thus may underestimate the significance.
Given that we even had to introduce the $\beta$ terms to achieve exactly this effect, this is not a problem.
With $2^{26}$ entries, we observed $>50\%$ of buckets to store a value $B[i]<0.1$, i.e., these buckets
do not contain frequent words or combinations.
This hashing trick has also been successfully applied to large textual streams
before in \cite{DBLP:conf/kdd/SchubertWK14,DBLP:conf/ssdbm/SchubertWK16} for the purpose of event detection,
and they report success with similar hash table sizes as we use.

\bigskip
In our first key contribution, we have defined a novel affinity probability, that captures how much
more likely words (co-) occur in a document compared to the background corpus. As second contribution,
we discussed how to modify t-SNE to work with this non-Euclidean affinity. Last, but not least important,
we propose a fixed-size memory sketch that provides a sufficient statistical summary of the word cooccurrence
frequencies in the background corpus to allow efficiently using this method with limited resources.

\section{Evaluation}\label{sec:eval}

For text processing we first use Stanford CoreNLP~\cite{DBLP:conf/acl/ManningSBFBM14}.
The main part for building the corpus sketch as well as the affinity computations is novel code.
For t-SNE layout and clustering, we build upon the code of the
open-source ELKI data mining framework~\cite{DBLP:journals/pvldb/SchubertKEZSZ15}, while
postprocessing and rendering are again new code.
Our corpus for normalization consists of all English Wikipedia articles as of 2017-05-01.
We use the Sweble Wikipedia parser to extract the text contents~\cite{DBLP:conf/wikis/DohrnR11}.
Since Wikipedia is a fairly large and diverse corpus, we can use a small bias constant
$\beta_C=1/|C|\approx 1.85 \cdot 10^{-7}$ (we used $|C|=5395344$ articles from the English Wikipedia 2017-05-01 dump).
For output rendering, we use JavaFX.

An objective evaluation of the resulting word clouds is difficult, and the optimal
solution depends strongly on the specific user goals~\cite{DBLP:conf/interact/LohmannZT09}.
Several metrics have been used in prior work, including
overlap, white space, size and stress \cite{DBLP:conf/smc/AdaTB10},
realized adjacencies, distortion, compactness, and uniform area utilization \cite{DBLP:conf/wea/BarthKP14}.
In comparison, 
however, it is fairly clear that measures such as overlap, white space and area utilization
are best optimized by other algorithms, and stress depends primarily on the similarity functions.
The number of realized adjacencies is likely the most adequate metric,
but the comparison methods expect a distance function rather than an odds
ratio or affinity probability, so we would be comparing apples and oranges.
Furthermore, we deliberately chose to apply the gravity heuristic because of visualization
aspects rather than best affinity projection, since the main use case for such word clouds
remains entertainment and explorative use. 
In the HCI community, an evaluation using user studies (e.g., \cite{DBLP:conf/iv/SeifertKKGG08})
and eye tracking (e.g., \cite{DBLP:conf/interact/DeutschST09}) would be common,
but we chose to not include initial results because of their bias:
as alternate word-clouds do not provide the word relationships, they obivously are
perceived to be less helpful for certain tasks; the remaining aspects are then entirely related to
the aesthetic point of view.
%
For evaluation, we discuss our individual
contributions, and contrast them
to alternatives, as we already did for the compression heuristic in \refsec{subsec:gravity}
with \reffig{fig:optimization}.

\subsection{Approximation Quality}

\begin{figure}[p]
\begin{subfigure}{\linewidth}\centering
\includegraphics[width=.8\linewidth]{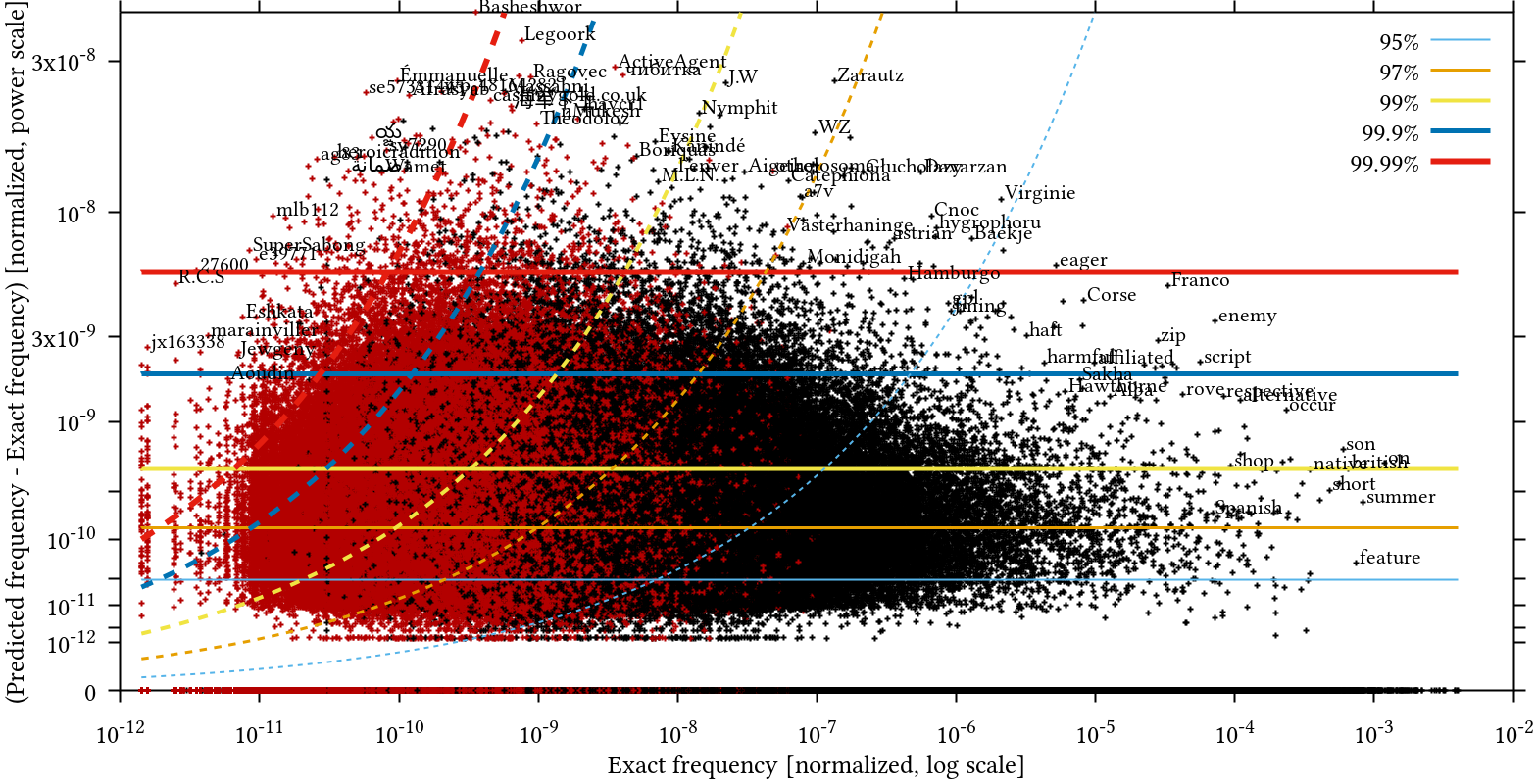}
\caption{Absolute error}
\end{subfigure}
\begin{subfigure}{\linewidth}\centering
\includegraphics[width=.8\linewidth]{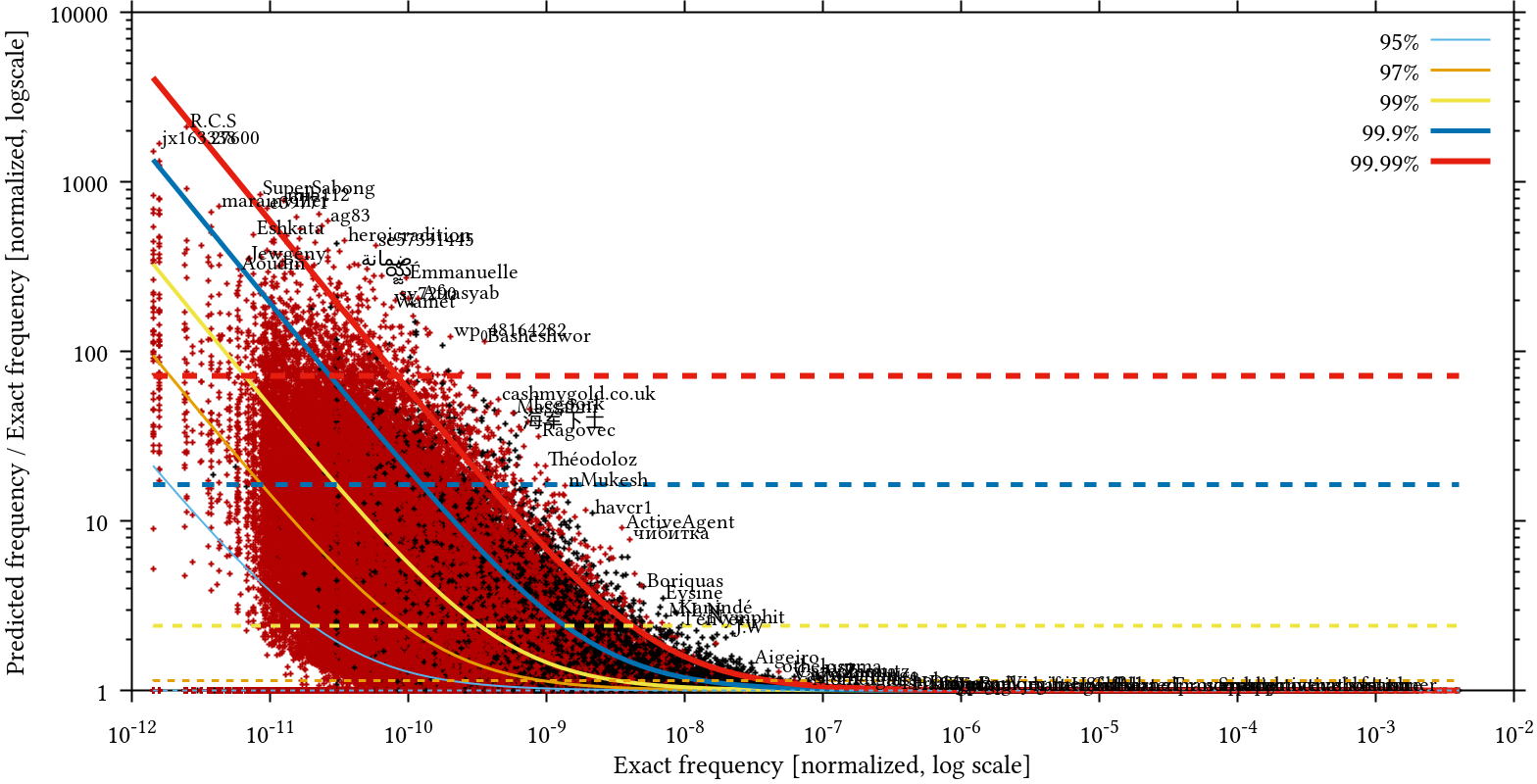}
\caption{Relative error}
\end{subfigure}
\begin{subfigure}{\linewidth}\centering
\includegraphics[width=.8\linewidth]{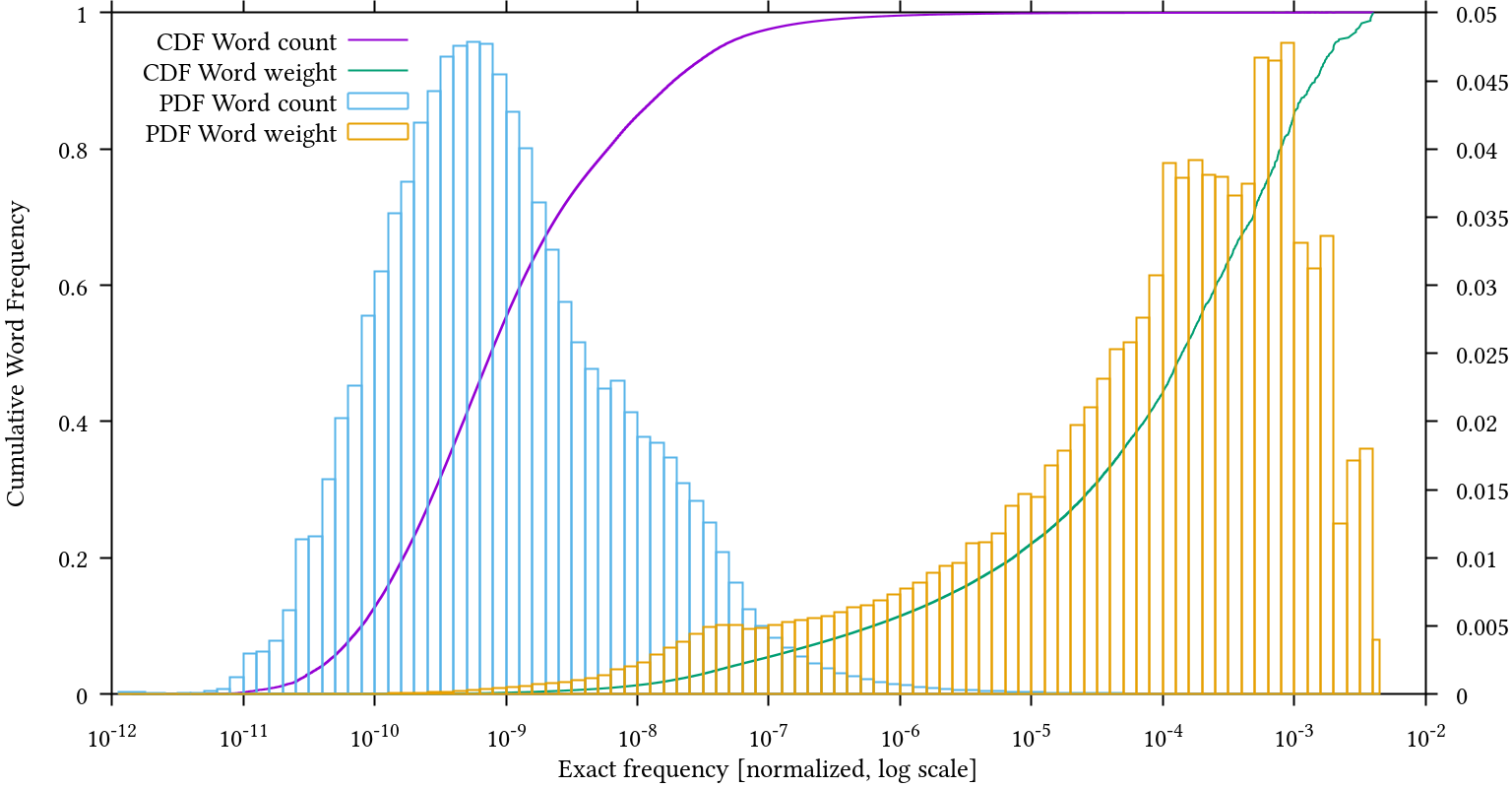}
\caption{Marginal distribution of word counts and word weight}\label{fig:approxerror-marg}
\end{subfigure}
\caption{Relative and absolute approximation errors. 
95\%, 97\%, 99\%, 99.9\%, respectively 99.99\%
of values are below the quantile lines (solid: absolute error, dashed: relative error),
for $>$93\% of the words, the errors is effectively zero.}
\label{fig:approxerror}
\end{figure}

First of all, we want to check if our corpus summary sketch (\refsec{subsec:approx}) provides
sufficiently accurate results, or if we make too many errors because of hash collisions.

In \reffig{fig:approxerror} we visualize the overestimation in relation to the term frequency,
and the most overestimated tokens. Red colors indicate words that occur in a single document only,
and account for 58\% of the words. For 93\% of all words, the error is exactly 0, and these points
are therefore on the x axis.
Prominent words with an unusually large absolute error are ``July'' and ``lesson''. However, the relative
error for these words is negligible ($0.0010\%$ resp.\ $0.0344\%$).
The largest overestimation happens for the word ``c.a.t.h.e.d.r.a.l.'', a song and album name occurring in
a single document only. Here, the true frequency is overestimated with $2.8\!\cdot\! 10^{-7}$ rather than $5.4\!\cdot\! 10^{-10}$.
The absolute error is negligible, and on the magnitude of our $\beta_C$ term.
If such errors are considered to be too severe, we could increase the number of buckets and hash functions. 
For example, if we increase the number of hash functions to $5$, ``c.a.t.h.e.d.r.a.l.'' no longer is overestimated.
In our experiments we found these errors to be tolerable; in particular since words can additionally be
detected because of cooccurrences.

\reffig{fig:approxerror-marg} shows the marginal distribution of the words,
once with every word having weight 1, and once using the average corpus frequency
$c_C(w)$ (\refeqn{eqn:weight-word}). We clearly see the long-tail effect here, that is, the majority of
words have a very low frequency (58\% of words occur in a single document only),
and a smaller set of words accounts for the majority of words.
Words with a frequency of less than $\approx 10^{-7}$ will have little impact on the result,
because of the $\beta_C$ term. In other words, rare words are considered to be random occurrences by our method.
The average frequency of all words is about $9.1\!\cdot\! 10^{-8}$, and the weighted average is about $4.8\!\cdot\! 10^{-4}$,
whereas the average absolute error is $2.3\!\cdot\! 10^{-11}$.

\subsection{Term Selection}

First of all, we want to verify our term selection method of \refsec{subsec:words}
by comparing the rankings of different methods, assuming that one uses a top-$k$ strategy
for selecting terms. We again use the ``Donald Trump'' Wikipedia article as example.
\reftab{tab:trump-rank} is sorted by the minimum rank of the different methods.
We include a simple frequency count ranking, a tf-idf ranking using idf from Wikipedia for normalization,
the single-word ranking per \refeqn{eqn:score-word}, and the ranking by taking both words
and word pairs into account as in \refeqn{eqn:score-max}.
For the latter two, we give the odds ratio (c.f.\ \refeqn{eqn:ratio}) and the
resulting probability (\refeqn{eqn:prob}).
We also experimented with the Brown corpus for \textrm{tf-idf} as used by~\cite{DBLP:conf/wea/BarthKP14},
but the results with the Wikipedia corpus were clearly better. 
The Brown corpus overly penalizes terms such as ``United States'' and ``president'' because it
contains many political news articles; but on the other hand it does not
contain e.g.\ the names ``Obama'' or ``Ivana'' (as the corpus is from the 1960s).
We can see that the methods yield surprisingly different results. For example, the term ``WrestleMania''
is only the 189th most frequent word in the article. With \textrm{tf-idf}, it is a top 50 word, and with
our new approach it is the top 12 relevant single word, but only gets a final rank of 209, because there are
many word cooccurrences with higher score. The terms ``United States'' and ``real estate'', for example,
are only selected due to taking cooccurrences into account.
The results when using only the word count 
are not convincing (it would select ``other'', ``time'' and ``year'' that are fairly common),
but \textrm{tf-idf} already does a reasonable job at selecting important words.
An interesting true positive is the word ``name'' as there even exists an article on Wikipedia titled
``List of Things named after Donald Trump''. While it is a common word, and thus ranked low by both
\textrm{tf-idf} and our single-word selection method, it scores higher because of cooccurrences with other words
~(supposedly with ``Trump'').

\begin{table}[tbp]\centering
\caption{Words from the ``Donald Trump'' Wikipedia article.\newline
Sorted by the minimum rank.
Bold indicates which method had the minimum rank, i.e., the strongest preference.
}
\label{tab:trump-rank}
{\fontsize{8}{8}\selectfont \setlength{\tabcolsep}{1.3pt}
\begin{tabular}{l|c|c|c|crr|crr}
Word                 & Min. & Count & tf-idf & \multicolumn{3}{c|}{Word Ratio} & \multicolumn{3}{c}{Pair Ratio} \\
                     & rank & rank  & rank   & rank & odds & prob. & rank & odds & prob. \\\hline
Trump                &   1 & \textbf{1} & \textbf{1} & \textbf{1} & 139.4 &  99.3 &   5 & 139.4 &  99.3 \\
casino               &   2 &  14 & \textbf{2} &  41 &   2.7 &  73.1 &  13 &  74.4 &  98.7 \\
say                  &   2 & \textbf{2} &   6 & 213 &   0.3 &  22.8 &   6 & 120.2 &  99.2 \\
York                 &   2 &   9 &  11 & 344 &   0.1 &  11.4 & \textbf{2} & 272.5 &  99.6 \\
Mar-a-Lago           &   2 & 303 &  31 & \textbf{2} &  45.2 &  97.8 &  35 &  45.2 &  97.8 \\
New                  &   2 &   4 &  25 & 607 &   0.0 &   4.4 & \textbf{2} & 272.5 &  99.6 \\
campaign             &   3 &   5 & \textbf{3} & 133 &   0.6 &  38.5 &   9 &  97.9 &  99.0 \\
goproud              &   3 & 454 &  55 & \textbf{3} &  33.2 &  97.1 &  63 &  33.2 &  97.1 \\
republican           &   4 &  10 & \textbf{4} & 149 &   0.6 &  35.5 &  34 &  46.2 &  97.9 \\
CPAC                 &   4 & 303 &  47 & \textbf{4} &  25.9 &  96.3 &  79 &  25.9 &  96.3 \\
States               &   4 &   9 &  18 & 431 &   0.1 &   7.8 & \textbf{4} & 249.8 &  99.6 \\
United               &   4 &   6 &  17 & 476 &   0.1 &   6.5 & \textbf{4} & 249.8 &  99.6 \\
state                &   4 & \textbf{4} &  14 & 499 &   0.1 &   6.1 &  17 &  61.4 &  98.4 \\
presidential         &   5 &  13 & \textbf{5} &  97 &   0.9 &  48.4 &  48 &  41.9 &  97.7 \\
Ivanka               &   5 & 454 &  77 & \textbf{5} &  15.1 &  93.8 & 144 &  15.1 &  93.8 \\
deferment            &   6 & 606 & 149 & \textbf{6} &  14.6 &  93.6 & 154 &  14.6 &  93.6 \\
Clinton              &   7 &  18 & \textbf{7} &  58 &   1.8 &  63.9 &  56 &  36.9 &  97.4 \\
estate               &   7 &  18 &  12 & 170 &   0.4 &  29.9 & \textbf{7} &  98.9 &  99.0 \\
real                 &   7 &  43 &  36 & 249 &   0.2 &  18.6 & \textbf{7} &  98.9 &  99.0 \\
President            &   7 & \textbf{7} &   9 & 287 &   0.2 &  14.8 &  51 &  39.1 &  97.5 \\
Trans-Pacific        &   7 & 454 &  95 & \textbf{7} &  14.3 &  93.5 &  98 &  21.9 &  95.6 \\
tax                  &   8 &  15 & \textbf{8} & 101 &   0.9 &  47.1 &  24 &  55.1 &  98.2 \\
trump-branded        &   8 & 909 & 158 & \textbf{8} &  13.0 &  92.8 & 184 &  13.0 &  92.8 \\
NYMA                 &   9 & 909 & 186 & \textbf{9} &  12.8 &  92.7 & 187 &  12.8 &  92.7 \\
sue                  &  10 &  29 & \textbf{10} &  59 &   1.8 &  63.8 &  11 &  93.2 &  98.9 \\
Organization         &  10 &  37 &  45 & 360 &   0.1 &  10.4 & \textbf{10} &  93.8 &  98.9 \\
Bethesda-by-the-Sea  &  10 & 909 & 173 & \textbf{10} &  12.4 &  92.5 & 193 &  12.4 &  92.5 \\
business             &  11 & \textbf{11} &  15 & 279 &   0.2 &  15.4 &  27 &  51.8 &  98.1 \\
make                 &  11 & \textbf{11} &  65 & 619 &   0.0 &   4.2 &  12 &  75.1 &  98.7 \\
alt-right            &  11 & 909 & 214 & \textbf{11} &  10.9 &  91.6 & 230 &  10.9 &  91.6 \\
WrestleMania         &  12 & 189 &  35 & \textbf{12} &  10.7 &  91.4 & 209 &  11.8 &  92.2 \\
Hotel                &  13 &  22 & \textbf{13} & 197 &   0.3 &  24.5 &  16 &  63.2 &  98.4 \\
non-interventionist  &  13 & 909 & 284 & \textbf{13} &  10.3 &  91.1 & 240 &  10.3 &  91.1 \\
hotel/casino         &  14 & 909 & 247 & \textbf{14} &  10.3 &  91.1 & 241 &  10.3 &  91.1 \\
golf                 &  15 &  46 &  19 & 172 &   0.4 &  29.7 & \textbf{15} &  63.3 &  98.4 \\
course               &  15 &  94 &  86 & 349 &   0.1 &  11.0 & \textbf{15} &  63.3 &  98.4 \\
hyperbole            &  15 & 606 & 175 & \textbf{15} &   9.6 &  90.6 & 260 &   9.6 &  90.6 \\
Fred                 &  16 &  37 & \textbf{16} & 127 &   0.7 &  40.0 &  32 &  49.3 &  98.0 \\
other                &  16 & \textbf{16} & 109 & 693 &   0.0 &   3.3 &  62 &  34.1 &  97.2 \\
Reince               &  16 & 909 & 250 & \textbf{16} &   9.3 &  90.3 & 270 &   9.3 &  90.3 \\
Trumped              &  17 & 909 & 230 & \textbf{17} &   9.2 &  90.2 & 274 &   9.2 &  90.2 \\
Party                &  18 & \textbf{18} &  26 & 492 &   0.1 &   6.2 &  34 &  46.2 &  97.9 \\
city                 &  18 &  27 &  79 & 731 &   0.0 &   2.9 & \textbf{18} &  58.8 &  98.3 \\
first                &  18 & \textbf{18} & 170 & 851 &   0.0 &   2.0 &  53 &  39.0 &  97.5 \\
Priebus              &  18 & 909 & 250 & \textbf{18} &   9.1 &  90.1 & 275 &   9.1 &  90.1 \\
become               &  19 &  26 & 112 & 671 &   0.0 &   3.5 & \textbf{19} &  57.9 &  98.3 \\
Lashley              &  19 & 606 & 150 & \textbf{19} &   7.7 &  88.4 & 305 &   7.7 &  88.4 \\
Ivana                &  20 & 151 & \textbf{20} & \textbf{20} &   6.9 &  87.3 & 244 &  10.1 &  91.0 \\
name                 &  20 &  29 & 141 & 859 &   0.0 &   2.0 & \textbf{20} &  57.4 &  98.3 \\
bankruptcy           &  21 &  75 & \textbf{21} &  60 &   1.8 &  63.8 & 462 &   5.6 &  84.9 \\
usfl                 &  21 & 303 &  63 & \textbf{21} &   6.8 &  87.2 & 377 &   6.8 &  87.2 \\
Obama                &  22 &  75 & \textbf{22} &  56 &   1.8 &  64.3 & 123 &  17.6 &  94.6 \\
Miss                 &  22 &  94 &  71 & 331 &   0.1 &  11.9 & \textbf{22} &  55.9 &  98.2 \\
show                 &  22 & \textbf{22} &  52 & 518 &   0.1 &   5.8 &  23 &  55.2 &  98.2 \\
Universe             &  22 & 227 & 143 & 244 &   0.2 &  19.0 & \textbf{22} &  55.9 &  98.2 \\
release              &  22 & \textbf{22} &  51 & 647 &   0.0 &   3.8 &  50 &  41.4 &  97.6 \\
time                 &  22 & \textbf{22} & 115 & 711 &   0.0 &   3.1 & 203 &  12.0 &  92.3 \\
year                 &  22 & \textbf{22} & 148 & 831 &   0.0 &   2.1 & 219 &  11.3 &  91.9 \\
Obamacare            &  22 & 909 & 325 & \textbf{22} &   6.7 &  87.0 & 385 &   6.7 &  87.0 \\
Donald               &  23 &  51 & \textbf{23} & 131 &   0.6 &  38.9 &  26 &  53.1 &  98.2 \\
Melania              &  23 & 606 & 136 & \textbf{23} &   6.4 &  86.6 & 403 &   6.4 &  86.6 \\
election             &  24 &  27 & \textbf{24} & 503 &   0.1 &   6.1 &  44 &  42.2 &  97.7 \\
return               &  24 &  75 & 147 & 556 &   0.1 &   5.3 & \textbf{24} &  55.1 &  98.2 \\
nbcuniversal         &  24 & 606 & 181 & \textbf{24} &   6.1 &  86.0 & 427 &   6.1 &  86.0 \\
Maples               &  25 & 454 &  97 & \textbf{25} &   6.1 &  85.9 & 428 &   6.1 &  85.9 \\
bondholder           &  26 & 909 & 333 & \textbf{26} &   5.8 &  85.2 & 449 &   5.8 &  85.2 \\
poll                 &  27 &  75 & \textbf{27} & 104 &   0.9 &  46.2 &  85 &  24.8 &  96.1 \\
Trumps               &  27 & 909 & 307 & \textbf{27} &   5.6 &  84.9 & 465 &   5.6 &  84.9 \\
NBC                  &  28 &  94 & \textbf{28} & 106 &   0.8 &  45.3 & 347 &   7.1 &  87.6 \\
Atlantic             &  28 & 170 & 105 & 303 &   0.2 &  13.8 & \textbf{28} &  51.2 &  98.1 \\
grope                &  28 & 909 & 349 & \textbf{28} &   4.8 &  82.8 & 502 &   4.8 &  82.8 \\
\end{tabular}}
\end{table}

\begin{figure}[tb]\centering
\begin{subfigure}[b]{.49\columnwidth}\centering
  \includegraphics[width=\linewidth]{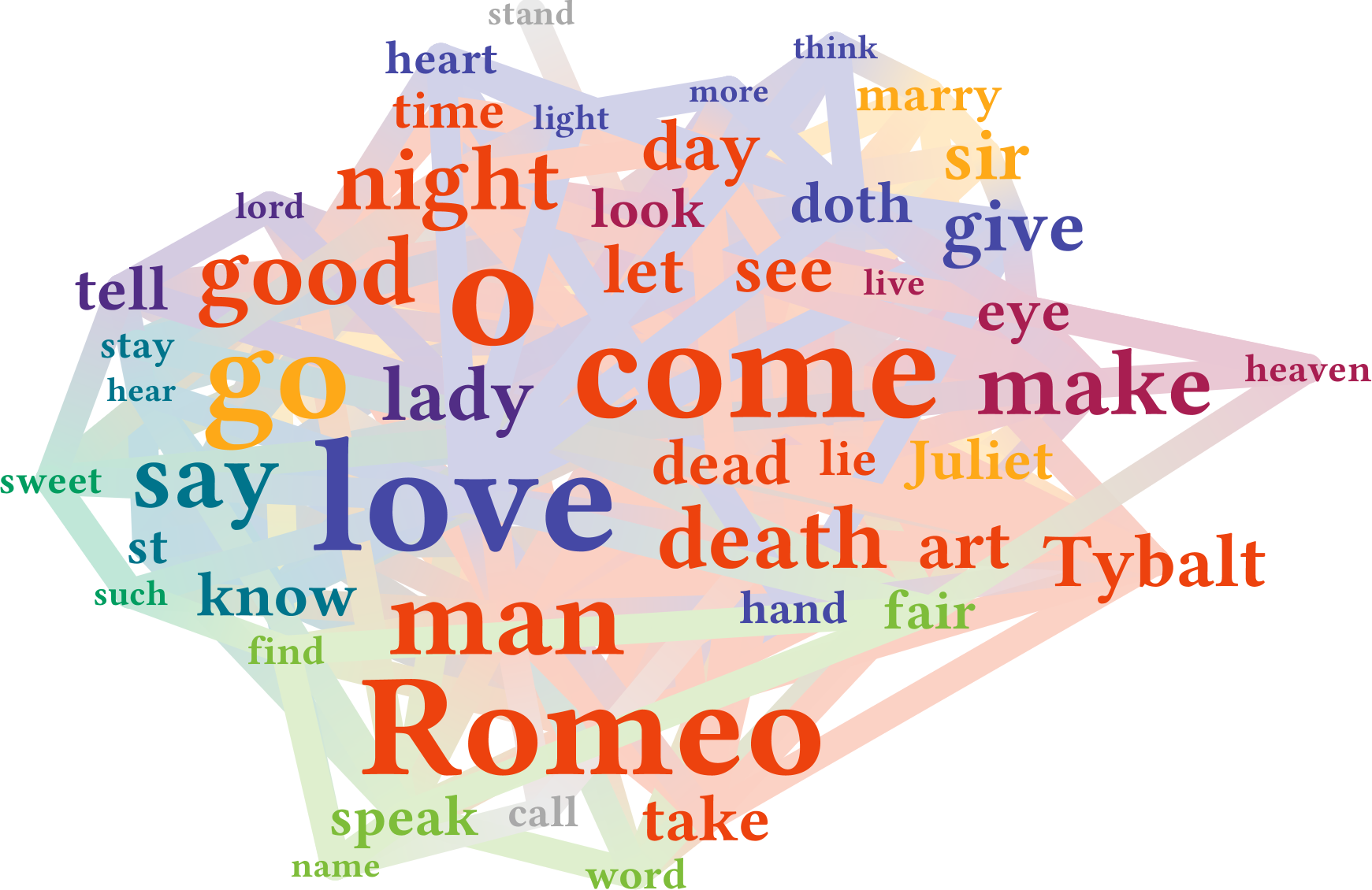}
  \caption{Constant prior word probabilities.}
  \label{fig:romeo-constant}
\end{subfigure}
\hfill
\begin{subfigure}[b]{.49\columnwidth}\centering
  \includegraphics[width=\linewidth]{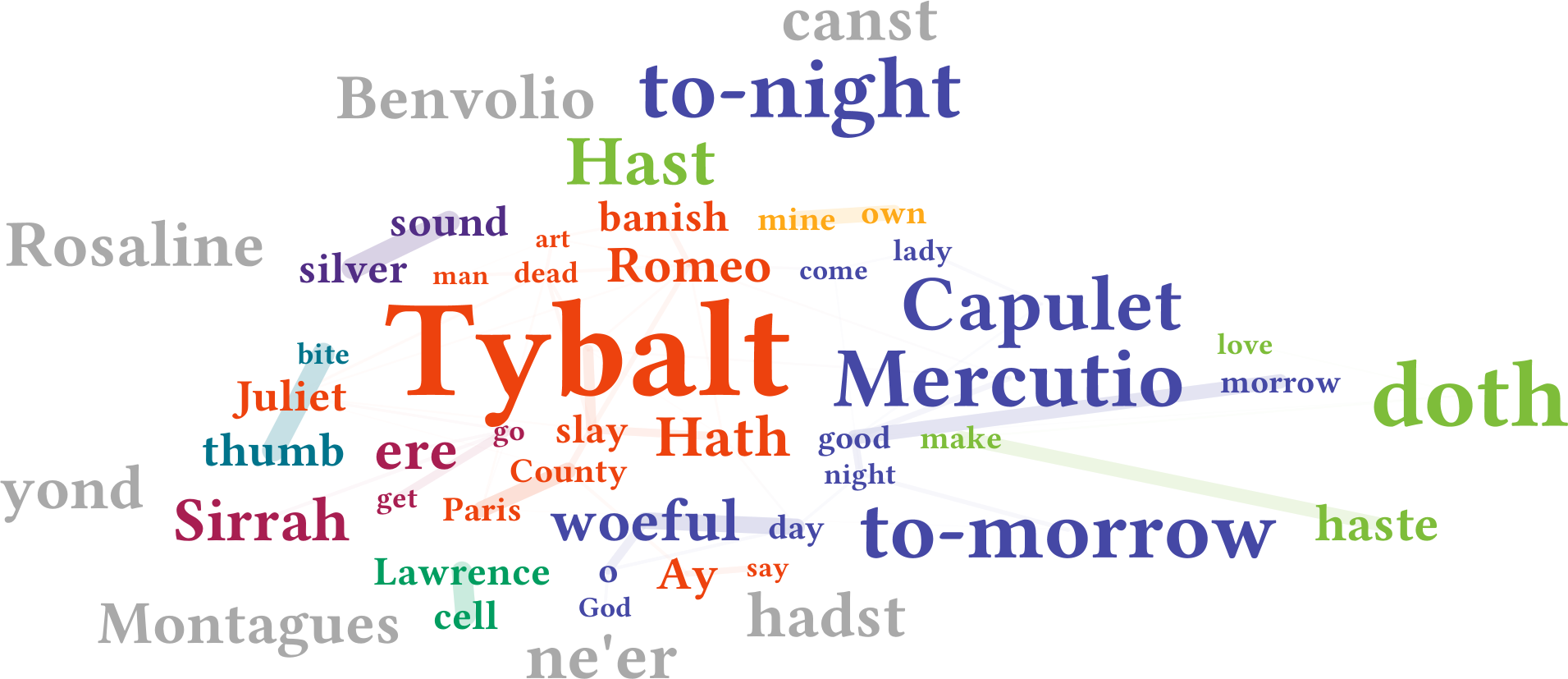}
  \caption{Wikipedia-based word normalization}
  \label{fig:romeo-wikipedia}
\end{subfigure}
\caption{Effect of preconditioning word frequencies on Shakespeare's play ``Romeo \& Juliet''.
Some words are significant because they are no longer in common use (e.g.\ ``doth''). 
With normalization, we pick up more characters 
and topics.
}
\label{fig:prior}
\end{figure}

\subsection{Corpus Normalization}

Next, we discuss the effects of using our background corpus (all English
Wikipedia articles) for normalization. As example text, we use Shakespeare's play ``Romeo and Juliet''
(note that \reffig{fig:optimization} used the Wikipedia article rather than the play).
In \refeqn{eqn:ratio}, we set the corpus cooccurrence frequencies $c_C=0$ to remove all influence of the background corpus on the result,
which causes our method to simply choose words based on their document frequency.
We show the resulting word cloud without using the corpus in \reffig{fig:romeo-constant},
while \reffig{fig:romeo-wikipedia} uses the Wikipedia corpus for normalization.
First of all, we lose the ability to
recognize ``significant'' cooccurrences, causing many more words to be connected such that the
data does not cluster well anymore. Some frequent words such as ``come'', ``go'' and ``make'' are
included that are probably not very characteristic and that would traditionally be removed as stopwords. 
Our new method does not require this filter since such words will have a low significance.
The less frequent, but important names ``Capulet'' and ``Montagues'' of the rivaling families are missing
in the pure frequency approach.
With corpus normalization, we additionally get e.g.\ the characters Mercutio, Benvolio, and Rosaline.
The ``silver sound'' of music is discussed at length by musicians at the end of Act IV.
But because the language of Shakespeare's play is very different from an average Wikipedia article,
we do have some words such as ``o'', ``doth'', ``hath'', and ``hast'' appear as unusual.
The scores of single words are often higher than those of word pairs in this example, so we do not get many connections.
These examples show the importance of choosing an appropriate corpus for normalization. Our Wikipedia corpus
is likely a good model of polished modern English; but when applied to a 16th century theater play
(even in a later edition; we used the Collins edition from Project Gutenberg) it is not entirely appropriate.
For specialized domains such as law texts, patent applications, medical reports, or summarization of
scientific publications, we expect better results if the corpus is chosen adequately.

\begin{figure*}[tbp]\centering
\begin{subfigure}[b]{.48\linewidth}\centering
  \includegraphics[width=.8\linewidth]{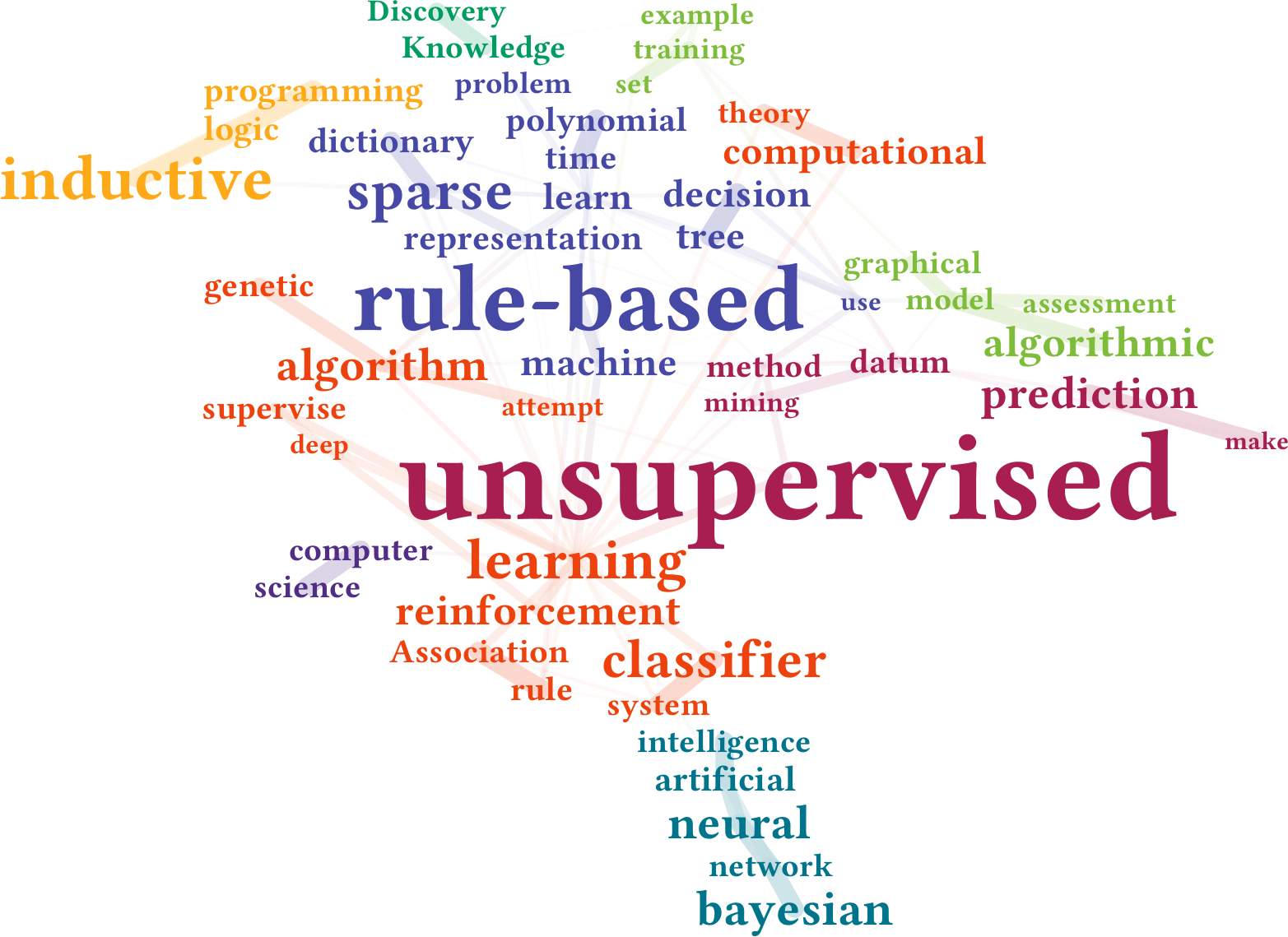}
  \caption{Random seed 0.}
  \label{fig:random0}
\end{subfigure}
\hfill
\begin{subfigure}[b]{.48\linewidth}\centering
  \includegraphics[width=\linewidth]{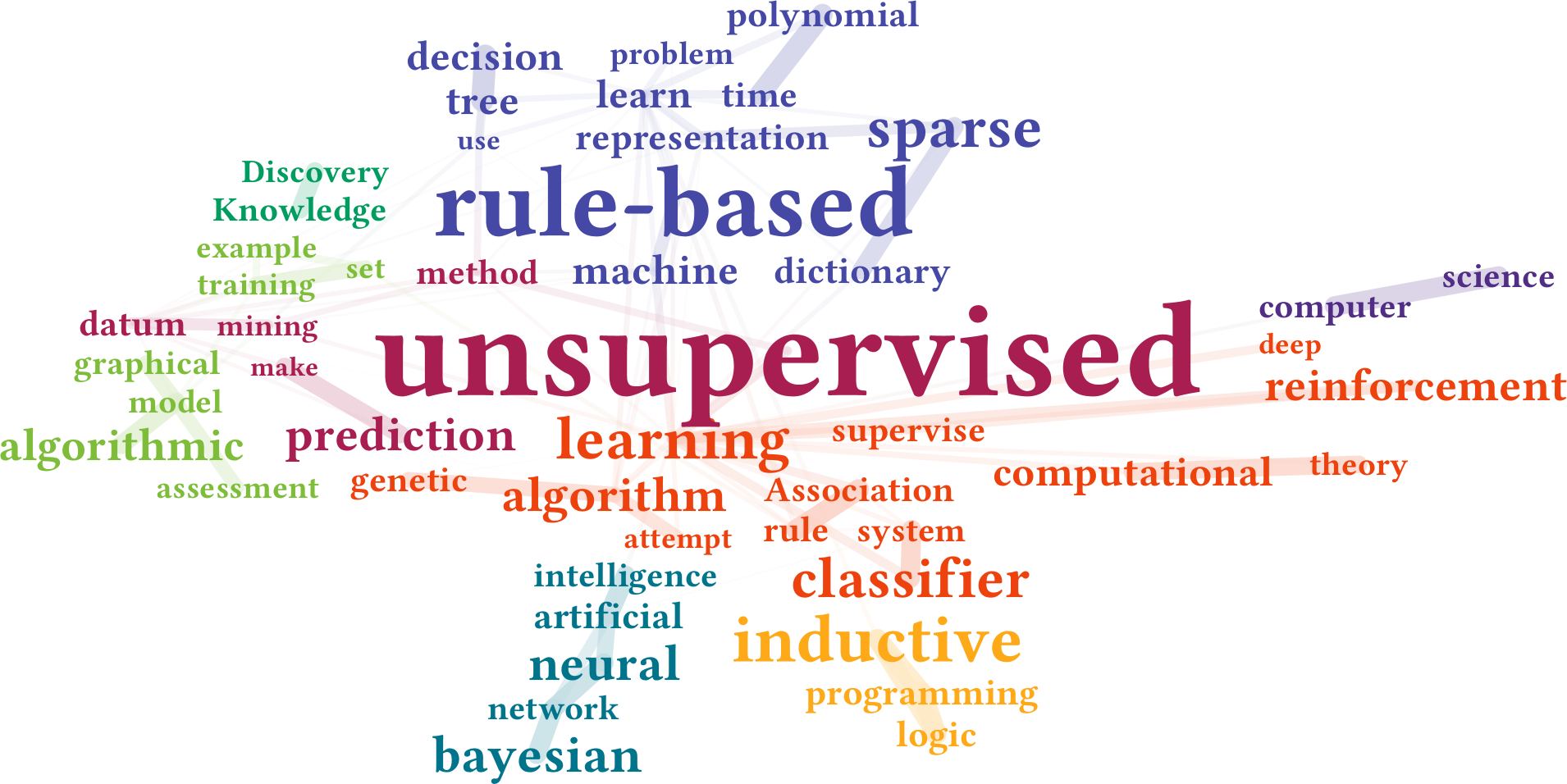}
  \caption{Random seed 1.}
  \label{fig:random1}
\end{subfigure}
\hfill
\begin{subfigure}[b]{.48\linewidth}\centering
  \includegraphics[width=\linewidth]{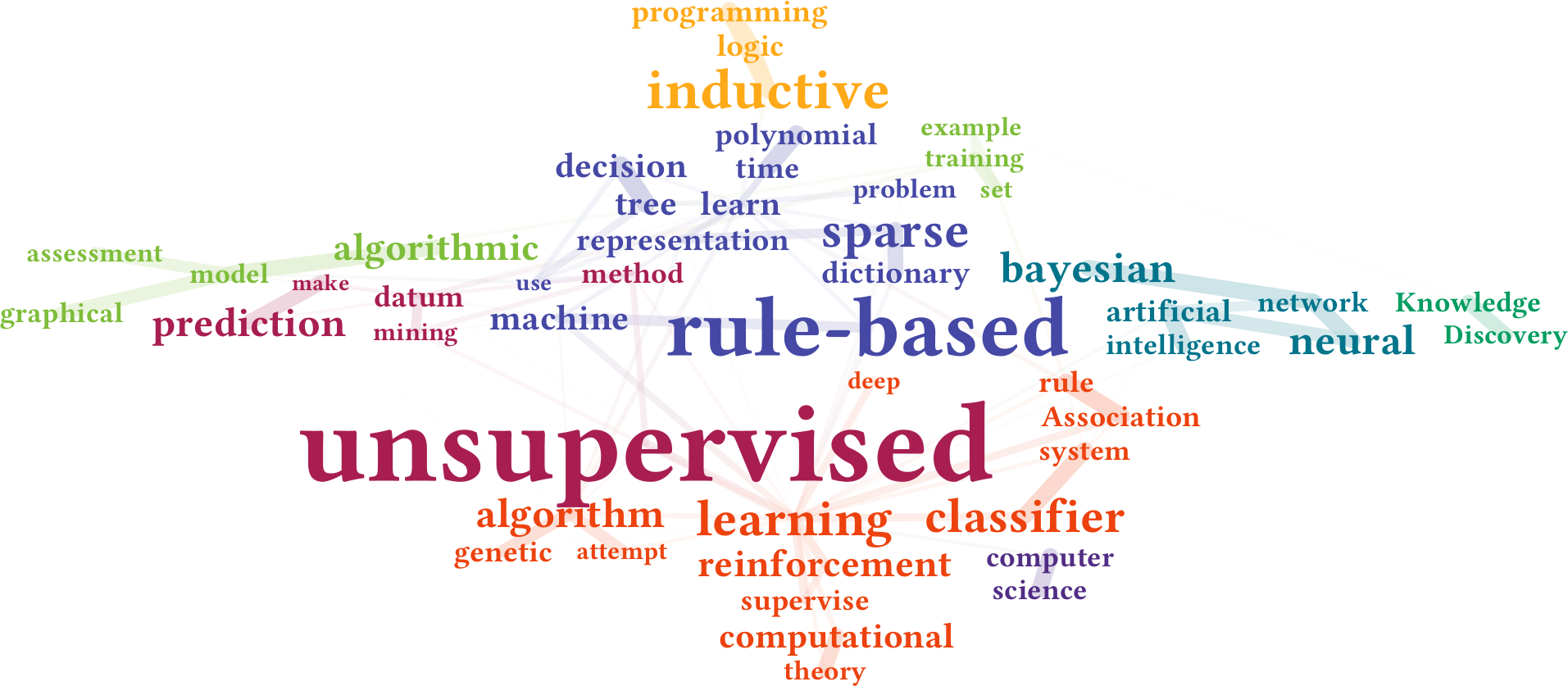}
  \caption{Random seed 2.}
  \label{fig:random2}
\end{subfigure}
\hfill
\begin{subfigure}[b]{.48\linewidth}\centering
  \includegraphics[width=\linewidth]{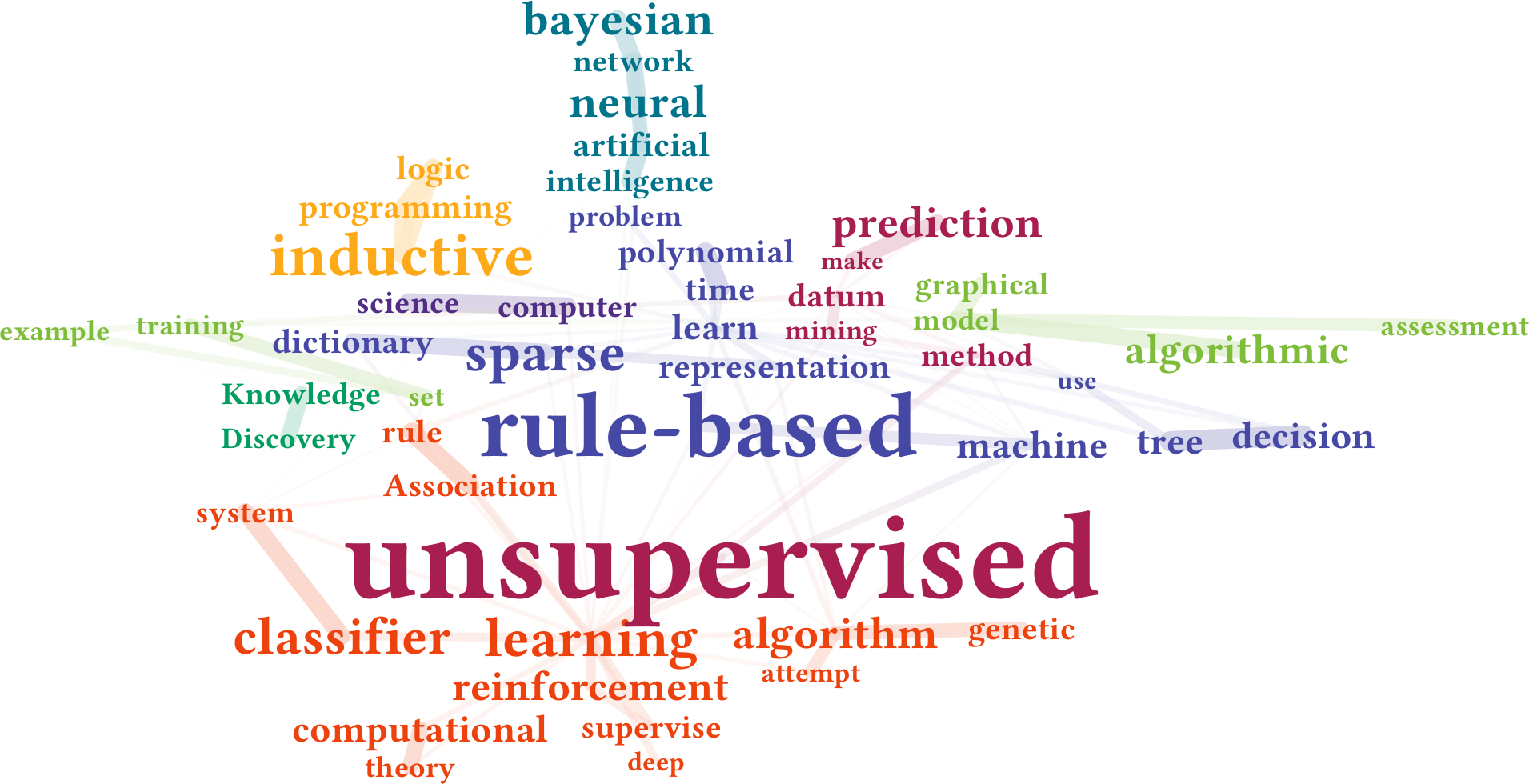}
  \caption{Random seed 3.}
  \label{fig:random3}
\end{subfigure}
\vspace{-1ex}
\caption{Effect of different random initializations on the visualization of the ``Machine learning'' Wikipedia article.
}
\label{fig:random}
\end{figure*}

\subsection{Stability and Randomness}

Given that t-SNE tries to minimize the Kullback-Leibler divergence,
one may be tempted to assume that the method produces rather
stable results. However, t-SNE starts its optimization procedure with a random
distribution in the output space. In this section we therefore want to test the
popular assumption that the results of t-SNE are rather stable except for rotation and
permutation of subgraphs. For this we use the Wikipedia article
``Machine learning'', and show the result using four different random seeds
in \reffig{fig:random} (for all other plots, we kept the random seed fixed to 0 for this paper).
Unfortunately, we find that the assumption does not hold as 
the results of the t-SNE algorithm vary much more than expected.
We attribute this to the many 0 affinities we have as opposed to the Euclidean-space
affinities traditionally used with t-SNE. The gravity
compression post-processing also partially contributes to the differences.
Because personal preferences vary, it is reasonable to allow the user to choose
the most pleasant plot from multiple runs. As noted by
Lohmann et al.~\cite{DBLP:conf/interact/LohmannZT09}, there is no single best arrangement,
and in their performance study, users partially even preferred layouts
that did \textit{not} deliver the best performance.

\section{Conclusions}\label{sec:conclusions}

In this work, we improved over existing methods for word cloud generation in several important aspects:

\begin{itemize}[leftmargin=*]
\item We introduce a word and cooccurrence significance score,
which uses word frequency relative to a large corpus such as Wikipedia instead of the absolute frequency.
\item We improve the keyword selection algorithm by using this score both on words and word cooccurrences.
\item We use t-SNE based on these affinities for improved word layout, and a compression algorithm to maximize word sizes.
\item We extract an improved clustering based on these affinities, and
we contribute a simple algorithm to cut the dendrogram tree to get a desired number of non-trivial clusters.
\item We include word relationships in the visualization to better represent the clustering structure of the cloud.
\end{itemize}

Nevertheless, there remains future work to be done.
The gravity compression algorithm could be replaced with more complex algorithms
such as seam-carving \cite{DBLP:journals/cgf/WuPWLM11} or
force transfer \cite{DBLP:conf/smc/AdaTB10}. Using the Delaunay triangulation as in \cite{DBLP:journals/cga/CuiWLWZQ10}
or moving subgraphs as a group could be used to retain the edge structure better when compressing the graph.
Such an optimization could be integrated into the t-SNE optimization process,
rather than being used in a second phase. This would allow a better preserving of affinities in the final layout,
while still providing readable font sizes without overlap.
We also did not employ pixel-exact optimization, which can save screen space, in particular for words with ascenders or descenders.
The new keyword selection method may also prove useful in other applications, which we also have not yet evaluated.

Last but not least, we would like to expand the presented approach to additional languages.
We also plan on providing a web service to allow users to easily try this approach on to their own texts
and web site contents
to maybe---to return to our running example---``make word clouds great again''.

\pagebreak
\bibliographystyle{ACM-Reference-Format}
\bibliography{literature} 

\end{document}